\documentclass[sigconf]{acmart}
\AtBeginDocument{%
  }

\copyrightyear{2025}
\acmYear{2025}
\setcopyright{cc}
\setcctype{by-sa}
\acmConference[UIST '25]{The 38th Annual ACM Symposium on User Interface Software and Technology}{September 28-October 1, 2025}{Busan, Republic of Korea}
\acmBooktitle{The 38th Annual ACM Symposium on User Interface Software and Technology (UIST '25), September 28-October 1, 2025, Busan, Republic of Korea}\acmDOI{10.1145/3746059.3747779}
\acmISBN{979-8-4007-2037-6/2025/09}

\usepackage{acmart-taps}
\usepackage{enumitem}
\usepackage{multirow}
\usepackage{graphicx}

\def\new{}

\makeatletter
\newcommand\newtag[2]{#1\def\@currentlabel{#1}\label{#2}}
\makeatother

\begin{document}

\title[Practical Guidelines for Biomechanical Simulations in HCI]{Demystifying Reward Design in Reinforcement Learning for Upper Extremity Interaction: Practical Guidelines for Biomechanical Simulations in HCI}

\author{Hannah Selder}
\orcid{0009-0008-7049-1630} 
\affiliation{%
  \institution{Center for Scalable Data Analytics and Artificial Intelligence (ScaDS.AI) Dresden/Leipzig, \\ Leipzig University}
  \city{Leipzig}
  \country{Germany}
}
\email{hannah.selder@uni-leipzig.de}

\author{Florian Fischer}
\orcid{0000-0001-7530-6838} 
\affiliation{%
  \department{Department of Engineering}
  \institution{University of Cambridge}
  \city{Cambridge}
  \country{United Kingdom}}
\email{fjf33@cam.ac.uk}

\author{Per Ola Kristensson}
\orcid{0000-0002-7139-871X} 
\affiliation{%
  \department{Department of Engineering}
  \institution{University of Cambridge}
  \city{Cambridge}
  \country{United Kingdom}}
\email{pok21@cam.ac.uk}

\author{Arthur Fleig}
\orcid{0000-0003-4987-7308} 
\affiliation{%
  \institution{Center for Scalable Data Analytics and Artificial Intelligence (ScaDS.AI) Dresden/Leipzig, \\ Leipzig University}
  \city{Leipzig}
  \country{Germany}
}
\email{arthur.fleig@uni-leipzig.de}

\begin{abstract} %

    Designing effective reward functions is critical for reinforcement learning-based biomechanical simulations, yet HCI researchers and practitioners often waste (computation) time with unintuitive trial-and-error tuning. This paper demystifies reward function design by systematically analyzing the impact of effort minimization, task completion bonuses, and target proximity incentives on typical HCI tasks such as pointing, tracking, and choice reaction. We show that proximity incentives are essential for guiding movement, while completion bonuses ensure task success. Effort terms, though optional, help refine motion regularity when appropriately scaled. We perform an extensive analysis of how sensitive task success and completion time depend on the weights of these three reward components. From these results we derive practical guidelines to create plausible biomechanical simulations without the need for reinforcement learning expertise, which we then validate on remote control and keyboard typing tasks. This paper advances simulation-based interaction design and evaluation in HCI by improving the efficiency and applicability of biomechanical user modeling for real-world interface development.
\end{abstract}

\begin{teaserfigure}
    \centering
    \includegraphics[width=\textwidth]{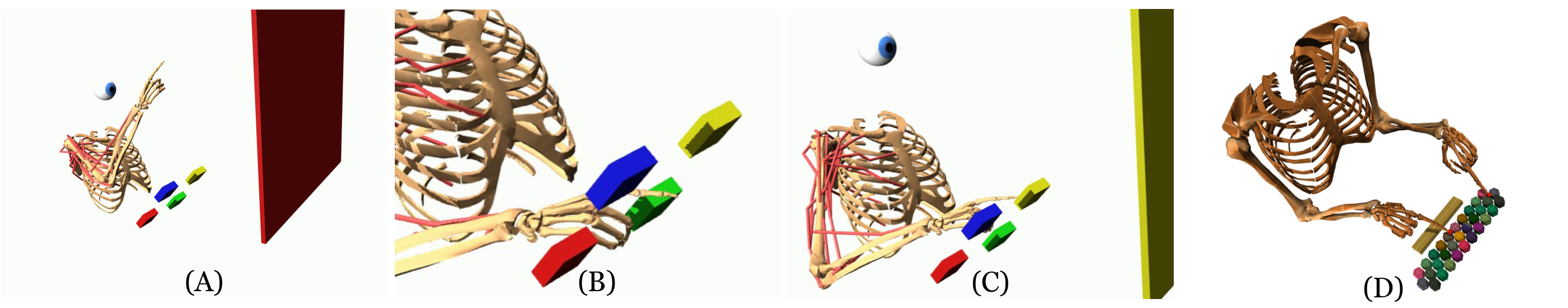}
    \caption{Training musculoskeletal user models via Reinforcement Learning can be a daunting task. We analyze reward function design to overcome typical failure modes, such as erratic movement without pressing the correct button in a choice reaction task (A) or moving towards the target button without pressing it (B). We distill guidelines on how to arrive at successful policies in the choice reaction task (C), bi-manual typing (D), and other interaction tasks (pointing, tracking, remote control).}
    \label{fig:teaser}
    \Description{Different movement patterns of human biomechanical models. (A), (B) and (C) show a model that is positioned in front of four buttons of different colors, with a display representing one of these colors. The agent represents an upper body, a right arm and one eye. In (A), the arm is directed upwards. In (B), the finger is positioned directly above one button but not pressing it. In (C), the finger is touching the button. The rightmost image (D) shows a biomechanical model with two arms pressing keys on a virtual keyboard.}
\end{teaserfigure}



\maketitle

\section{Introduction}

Reinforcement Learning (RL) has emerged as a transformative force across a wide range of domains, including robotics~\cite{figureai2025,doi:10.1126/scirobotics.adi8022}, gaming~\cite{simateam2024scalinginstructableagentssimulated}, and autonomous systems~\cite{WU2024104654}.
In Human-Computer Interaction (HCI), RL is increasingly used to drive biomechanical user simulations that model plausible human behavior during interaction~\cite{Hetzel2021,fischer2021reinforcement,ikkala_breathing_2022}, opening the door to a dynamic, embodied understanding of movement-based tasks and virtual prototyping.
Like moths to a flame, researchers are drawn to these powerful tools -- only to discover that the flame might burn them too: polished results often come at the cost of extensive, unintuitive trial-and-error.

At the heart of RL lies the \textit{reward function}, a mathematical description of the agent’s goals, such as typing a word on a keyboard (Figure~\ref{fig:teaser}).
This function encodes what it means to act successfully and plausibly: 
For biomechanical simulations, did the agent, i.e., the biomechanical user model, reach the correct target? 
Was the motion biomechanically plausible? 
While prior works have shown that RL-trained agents can simulate plausible user behavior in HCI~\cite{fischer2021reinforcement, ikkala_breathing_2022}, they rarely reveal how much time and effort went into tuning these reward functions. 
For state-of-the-art %
musculoskeletal models, each design tweak %
can take 12–72 hours on modern workstations to train -- potentially resulting in subtle, non-intuitive differences. 
A lack of general guidance on reward function design, limited understanding of how different components affect outcomes, and high computational costs of unstructured trial-and-error tuning all constitute a major entry barrier for HCI researchers without RL expertise.

\new{Extending preliminary results~\cite{selder25chilbw},} this paper aims to demystify reward function design and provide practical guidance %
that lowers barriers for newcomers while refining practice for experts. 
\new{While individual reward terms may seem reasonable, combining them unsystematically can cause agents to exploit loopholes and fail the task. Band-aid fixes, like adding penalty terms for specific unwanted behaviors, tend to result in overly complex and “overfitted” rewards. This problem is especially pronounced in high-dimensional, over-actuated musculoskeletal models and black-box dynamics. %
To tackle this issue,%
}
we systematically analyze how reward components -- target proximity, task completion bonuses, and effort minimization -- impact learning outcomes across movement-based HCI tasks such as pointing, tracking, and choice reaction. 
We look at how sensitive these outcomes are to changes in the weights of the individual reward components. 
In total, we trained and tested over 500 policies -- so other researchers don’t have to. 
From this, we derive practical, \new{sequential} guidelines to accelerate and simplify biomechanical simulation design. 
\new{For instance, finding that effort terms are not always necessary to generate successful movement trajectories, we suggest to omit them unless movement instabilities such as trembling arise.}
We validate these guidelines on two more advanced HCI tasks that do not only involve aimed arm movements, but also require learning to control a physical input device (joystick and keyboard).

While our goal is to support more principled and accessible reward design, some limitations are inherent to our scope. 
Our analysis focuses on basic visuomotor interaction tasks that do not require much creative or strategic thinking.
Tasks that involve %
highly ambiguous goal structures or leave room for a diversity of creative strategies may require complementary methods such as hierarchical RL and more sophisticated cognitive modeling. 
Similarly, our evaluation is currently grounded in arm-based interaction. %

In summary, we address the following \textbf{research questions}: 
\aptLtoX{\begin{enumerate}
    \item[\textbf{RQ.1}]\label{item:rq:plausible-movement}\textbf{(Plausibility)} What combinations and relative weightings of reward function components (i.e., proximity, effort, and task completion bonuses) produce plausible human movement trajectories across a range of goal-directed visuomotor tasks (pointing, tracking, choice reaction)?
    \item[\textbf{RQ.2}]\label{item:rq:sensititivies} \textbf{(Sensitivity)} What are the sensitivities of interaction outcomes (e.g., completion time, success rate) to variations in individual reward components in these tasks?
    \item[\textbf{RQ.3}]\label{item:rq:generalisabiliy} \textbf{(Generalizability)} To what extent do the observed findings on the effects of reward components generalize across a set of goal-directed visuomotor tasks?
 	\end{enumerate}}{
\begin{enumerate}[label=\textbf{RQ.\arabic*}]
    \item\label{item:rq:plausible-movement} \textbf{(Plausibility)} What combinations and relative weightings of reward function components (i.e., proximity, effort, and task completion bonuses) produce plausible human movement trajectories across a range of goal-directed visuomotor tasks (pointing, tracking, choice reaction)?
    \item\label{item:rq:sensititivies} \textbf{(Sensitivity)} What are the sensitivities of interaction outcomes (e.g., completion time, success rate) to variations in individual reward components in these tasks?
    \item\label{item:rq:generalisabiliy} \textbf{(Generalizability)} To what extent do the observed findings on the effects of reward components generalize across a set of goal-directed visuomotor tasks? 
\end{enumerate}}
With \ref{item:rq:plausible-movement}, we aim to support the reward function design process by identifying combinations that lead to plausible simulated behavior.
With \ref{item:rq:sensititivies}, we investigate how sensitive key performance metrics are to individual reward components, helping researchers anticipate the effects of design changes.
With \ref{item:rq:generalisabiliy}, we seek to uncover recurring effects and trade-offs across multiple tasks, enabling the formulation of general reward function design guidelines.

For these guidelines to be useful, we believe they should exhibit the following key qualities: they should be (i) \textbf{succinct} to reduce cognitive load; (ii) \textbf{clear} in what each point contributes; (iii) \textbf{sequential} to offer a logical flow that eases step-by-step implementation; (iv) \textbf{actionable} with concrete steps or heuristics; and (v) \textbf{justified} by empirical evidence or clear reasoning. 
These principles informed both our analysis and the way we present our findings.

We \textbf{contribute} to the computational modeling of movement-based interaction by: 
\begin{itemize} 
    \item presenting the first systematic and empirical analysis of reward function components (proximity, effort, and task bonuses) across three canonical goal-directed HCI tasks (pointing, tracking, choice reaction); 
    \item revealing key sensitivities and trade-offs that emerge during reward function tuning, thus deepening the understanding of RL behavior in HCI contexts;  
    \item providing clear, actionable, and empirically grounded guidelines for reward function design that lower the entry barrier for newcomers and support experts.
\end{itemize}

\new{To support reproducibility and facilitate further research, we release a dataset comprising 517 policies trained for the reward functions discussed in this paper~\cite{dataset}. These policies are intended for use within the User-in-the-Box framework \cite{ikkala_breathing_2022}. The dataset can be found at \url{https://doi.org/10.5281/zenodo.15845404}.
}

\section{Related Work}
Our goal in this section is twofold: (i) to contextualize our contribution within the growing body of work on RL-based biomechanical simulations, and (ii) to offer an entry point for HCI researchers, especially those without prior RL expertise, who are interested in leveraging these methods for interaction design. 
To this end, we survey how biomechanical models have been applied in HCI to simulate, prototype, and evaluate movement-based interactions in Section~\ref{subsec:biomechanical-simulations-in-hci}. 
We then discuss the role of reward function design in RL and its implications for plausible agent behavior in Section~\ref{subsec:reward-fct-design}, before summarizing the identified research gaps in Section~\ref{subsec:research-gaps}.
\new{
While extensive literature has developed guidelines and workflows supporting different aspects of developing and testing user simulation, including computational models of human behavior and rationality~\cite{chandramouli_workflow_nodate, zhang_using_2020}, empirical design decisions~\cite{patterson_empirical_2024}, and integrating behavioral data into model formulation~\cite{wilson_ten_2019}, this work addresses practical aspects in RL-based user modeling, namely the development and iterative refinement of reward functions. }

\subsection{Biomechanical Simulations in HCI}\label{subsec:biomechanical-simulations-in-hci}

\new{Simulation-based approaches are gaining increasing attention in HCI research, particularly for modeling user behavior during interaction. Recent work highlights the relevance of RL-based user modeling, including simulations of touchscreen typing~\cite{shi_crtypist_2024} and adaptive interfaces supported by learned user models~\cite{langerak_marlui_2024}. In parallel,} biomechanical simulations have evolved as a beneficial tool for developing and validating HCI technologies~\cite{murray-smith_what_2022, bachynskyi2015performance}.
They are fundamentally based on kinematic and dynamic models, which, depending on their complexity, provide information on the skeletal structure, inertial properties, and the neuromuscular system of the human body~\cite{lee2009comprehensive}. 
While early models were limited to calculating mechanical loads in static postures~\cite{winter_biomechanics_1984, flash_coordination_1985, Ayoub_Biomechanical_1974}, advancements have led to physiologically more accurate musculoskeletal models~\cite{madeleine_biomechanics_2011, holzbaur2005model, damsgaard_analysis_2006, delp_opensim_2007, lee2009comprehensive}, which can be used to predict ergonomics, fatigue, and user strategies in addition to traditional performance metrics~\cite{fischer_sim2vr_2024}.
Biomechanical simulations therefore provide valuable insights into predictability, safety, and accessibility, making them a promising approach for system evaluation prior to and as a complement to user testing \cite{murray-smith_what_2022}.

Traditionally, biomechanical models integrated into physics engines such as OpenSim~\cite{delp_opensim_2007} have been used for \textit{inverse simulation}. Inverse methods aim to infer dynamic properties and motor patterns from reference motion data, enabling in-depth analysis of the physiological state and the development of assistive tools, for example for rehabilitation~\cite{koopman1995inverse, pizzolato2017real}. 
In the context of virtual reality, \citeauthor{hwang_ergopulse_2024} apply simulations with electrical muscle stimulation to create kinesthetic force feedback for the lower body~\cite{hwang_ergopulse_2024}.

A second and more recent branch of research uses \textit{forward simulation} to predict human movements and ergonomic states from muscle control signals~\cite{ackermann2010optimality, maas2013biomechanical}.
These methods require defining a controller that actively selects neural muscle signals based on the sensed body state and the given interaction task.
The framework of optimal feedback control has emerged as the de facto standard to model how these muscle signals are selected in aimed movements~\cite{todorov_optimal_2002, fischer_optimal_2022}.
Optimal feedback control is based on the assumption that information on the controlled quantity (e.g., the arm, hand, and controller in VR interaction) is continuously fed back to the controller, resulting in a closed interaction loop between the user and their environment.
Optimal feedback control methods such as LQG~\cite{fischer_optimal_2022}, MPC~\cite{klar_simulating_2023}, and intermittent control~\cite{martin_intermittent_2021} have been successfully applied to predict movement trajectories in standard interaction tasks such as mouse pointing and mid-air pointing.
However, they impose severe restrictions on the complexity of the biomechanical models and tasks considered. For example, simplified activation models based on torque actuators directly attached to the joints have been introduced~\cite{klar_simulating_2023, Hetzel2021} to overcome the complexities arising from the highly over-actuated, noisy, and delayed neuromuscular system~\cite{Leib24}.

In recent years, deep RL has emerged as the go-to method for forward simulations of movement-based interaction~\cite{song2021deep, caggiano2022myosuite}.
Building on the theory of optimal feedback control, deep RL methods approximate the optimal control policy through efficient exploration and learning from interaction with an unknown environment~\cite{sutton1998reinforcement}. In contrast to traditional methods, RL methods do not require a closed formulation of the system to be controlled, which makes them particularly useful for non-linear musculoskeletal systems and black-box physics engines such as OpenSim~\cite{delp_opensim_2007} or MuJoCo~\cite{todorov_mujoco_12}.
For example, \citeauthor{fischer2021reinforcement} use SAC, an entropy-based RL method, to learn controlling the muscles of a state-of-the-art shoulder model for a mid-air pointing task.
\citeauthor{Hetzel2021} apply deep RL to simulate arm movement in mid-air typing.

The combination of motor and sensory control models, as discussed in \cite{nakada_deep_2018}, has been particularly relevant for modeling human-computer interaction. %
\citeauthor{ikkala_breathing_2022} present \textit{User-in-the-Box}, an RL-based simulation framework to generate task-specific movement trajectories based on the user's visual and proprioceptive perception of the interaction environment~\cite{ikkala_breathing_2022}. %
\citeauthor{moon_real-time_2024} use movement data from biomechanical simulations to infer aimed target positions via neural density estimation~\cite{moon_real-time_2024}.
\citeauthor{fischer_sim2vr_2024} develop a system for running biomechanical models directly in VR applications, ensuring that the simulated user "perceives and controls" exactly the same environment as a real user~\cite{fischer_sim2vr_2024}.
With recent improvements to training efficiency, the scope of RL-based simulations has been extended to dexterous manipulation and grasping tasks~\cite{schumacher_dep-rl_2022, chiappa2023latent, caggiano2023myodex, berg2024sar}.

\subsection{Reward Function Design for RL-Based Simulations}\label{subsec:reward-fct-design}
In RL-based biomechanical simulations, the design of the reward function is identified as a key factor in the effectiveness of the learning process \cite{kwiatkowski_reward_2023}.
This is because most RL methods (with a few noteworthy exceptions~\cite{rajeswaran2017learning, peng2018deepmimic}) do not learn from human reference data but solely from a predefined reward function. 
While different formulations of effective reward functions have been examined \cite{kwiatkowski_reward_2023, he_exploring_2024, nowakowski_human_2021,ikkala_breathing_2022}, most of them have been handcrafted for a specific task~\cite{caggiano2023myodex, caggiano2022myosuite, ikkala_breathing_2022}, which limits their generalizability across tasks and contexts. 
\new{In this process, even if a "working" reward function is found, it is often unclear to what extent it accurately encodes the preferences of a human stakeholder~\cite{muslimani_towards_2025}.}
In addition, there is a lack of benchmark tasks and environments that would enable a fair comparison between proposed reward functions. %
Instead, the common practice is to identify and validate a reward function based on trial-and-error within a unique, task-specific setting that contains a large number of potentially confounding variables (and is often difficult to replicate). These confounding factors may include the biomechanical model used, different types of control and exploration noise~\cite{charaja_generating_2024}, as well as factors modulating the training process such as early termination constraints~\cite{kumar2021modelling, ikkala_breathing_2022}, predefined or adaptive learning curricula~\cite{fischer2021reinforcement, Hetzel2021, fischer_sim2vr_2024}, and even the RL algorithm used~\cite{booth_perils_2023}. %

In addition, %
movement-based interaction
usually involves a trade-off between two or more opposing objectives, e.g., between accuracy and stability~\cite{liu2007evidence} or speed and accuracy~\cite{nagengast2011risk}. 
While for many practical tasks it is natural to only define \textit{sparse rewards} (e.g., providing a single positive reward if and only if the desired target position or velocity has been reached~\cite{fischer2021reinforcement, charaja_generating_2024}, or when the game score increases~\cite{mnih2013playing, fischer_sim2vr_2024}), this sparsity of information leads to very slow training and often hinders convergence.
As a consequence, RL literature has explored adding continuous \textit{dense} reward terms that provide information about the utility of a state-action pair much earlier during an episode (e.g., moving closer towards vs.\ moving away from the desired target position).
In particular, the theory of \textit{reward shaping} has been developed~\cite{ng1999policy}, and bi-level approaches have been proposed to increase the robustness of RL training while adhering to the goals of the original sparse reward formulation~\cite{devidze_explicable_nodate, hu2020learning}. %
However, these approaches are rarely used in practice, probably because of their complexity and the lack of toolkits implementing them. 
Instead, ad-hoc approaches based on trial-and-error are predominate~\cite{booth_perils_2023, nowakowski_human_2021}. %
Since 'failed' reward functions explored along the way are rarely reported, little insight is gained into which reward patterns are critical for successful RL training, essentially leaving all RL designers with the same difficulties.
\new{In this work, %
we propose an intermediate approach that systematically adds task-relevant 'guiding' terms in order to shape the reward function without introducing systematic bias, although the latter can only be experimentally validated.}

The complexity of designing an appropriate reward function is further exacerbated in biomechanical simulations, since humans can perform tasks with an infinite number of different admissible joint trajectories and muscle activations~\cite{berret_evidence_2011} ("movement redundancy"~\cite{neilson1993problem}).
To limit the possibilities to "sensible" movements, at least one \textit{effort} term is typically added for regularization, e.g., to restrict the use of rapid and abrupt arm movements by penalizing large muscle activations, thus ensuring that available resources are used efficiently.
Several effort cost models, including jerk~\cite{flash_coordination_1985}, energy consumption~\cite{nelson1983physical}, and commanded torque change~\cite{nakano1999quantitative}, have been proposed and investigated from a motor control perspective~\cite{berret_evidence_2011, wang_advances_2011, cheema_predicting_2020, charaja_generating_2024}.
The role of an effort model in motor adaptation has been explored in empirical studies in~\cite{xu_inducing_2021, proietti_modifying_2017}.
While there is consensus in the literature that effort reduction is likely to play a key role in movement planning, it is unclear how this incentive can be reconciled with other task- or performance-related goals~\cite{Leib24}. 

\subsection{Identified Research Gaps}\label{subsec:research-gaps}
To the best of our knowledge, there exist no guidelines on how to design and balance the individual reward components in practical HCI tasks, especially in combination with complex musculoskeletal systems.
From an engineering point of view, it is unclear to what extent the different effort models can help guide the RL training towards biomechanically plausible movements and %
improve convergence in RL-based simulations.
We therefore anticipate a strong need to explore the design of reward functions for realistic use cases of biomechanical models. 
In this work, we take a decisive step towards this goal by starting with a number of goal-directed visuomotor tasks that are highly relevant for HCI and analyzing the individual and combined effects of different reward function components on RL-based learning of interactive body movements.

\section{Methodology}
In this work, we systematically analyze how different reward components (which we introduce in Section~\ref{subsec:reward-components}) influence emergent user strategies in RL-based biomechanical simulations. 
Our focus lies on visuomotor interaction tasks involving direct input via the index finger, such as pointing, tracking, and typing. 
Section~\ref{sec:tasks} outlines the full range of tasks we consider, while Section~\ref{subsec:exp-design} details our experimental design and rationale.

\subsection{Reward Components}\label{subsec:reward-components}
In designing the reward function, we focus on three components:
\begin{itemize}
    \item %

    The \textit{task bonus} is a term that incentivizes reaching states that are beneficial for successfully completing the given task. Often an episode terminates immediately after such a state is reached. %
    We integrate the many possibilities in the function $f_{\text{bonus}}(\cdot)$, where $(\cdot)$ is a placeholder for all relevant function arguments.
    Defining such a bonus %
    only requires splitting the state space into a set of desirable "goal states" to be awarded (e.g., when a %
    desired button is hit), and a set of neutral states, where no information regarding their desirability is provided.
    
    \item The \textit{distance} term provides a proximity incentive by rewarding the agent more the closer they get to the target, expressed by 
    the function $f_{\text{distance}}(\cdot)$.
    \item 
    The \textit{effort} model aims to incentivize controls that require low physiological effort. Depending on which measure of effort is used, this model can be quite versatile:
    Designers can choose, e.g., to penalize jerky movements, or reward movements that require lower energy. 
    We encompass the possibilities in the function $f_{\text{effort}}(\cdot)$. 
\end{itemize}
To evaluate the intricacies of how the individual components work independently and in conjunction, for each component we introduce respective weights $w_{\text{distance}}, w_{\text{effort}}, w_{\text{bonus}} \geq 0$. 
In total, the most generic reward function amounts to 
\begin{equation}\label{eq:reward-fct-composite}
    r_t = w_{\text{bonus}} \cdot f_{\text{bonus}}(\cdot) - w_{\text{distance}} \cdot f_{\text{distance}}(\cdot) - w_{\text{effort}} \cdot f_{\text{effort}}(\cdot).
\end{equation}
If we set $w_{\text{effort}}=0$ or $w_{\text{distance}}=0$, we speak of \textit{zero effort} or \textit{zero distance}, respectively.

The \textbf{task bonus} is formulated as follows:
\begin{equation}\label{eq:bonus}
    f_\text{bonus}(\cdot) = \begin{cases}
        1, & \text{if model reaches a goal state, } \\
        0& \text{else}.
    \end{cases}\tag{Bonus}
\end{equation}

We investigate three different \textbf{distance} reward functions, each based on the distance between the index finger and the task-specific target position (either a point or a surface), $dist$, as measured by a MuJoCo distance sensor~\cite{todorov_mujoco_12}. In every timestep, we can measure:

\begin{enumerate}
    \item The (absolute) value of the MuJoCo distance sensor: 
\end{enumerate}

    \begin{equation*} \label{eq:absolute_dist}
        f_{\text{distance}}(dist) = \|dist\|_{2}  \tag{$D_\text{absolute}$}
    \end{equation*}
\begin{enumerate}
    \item[(2)] The squared distance, which has been successfully used in RL tasks~\cite{nocedal_numerical__2006, kim_approach_2023}:%
\end{enumerate}
    \begin{equation*} \label{eq:squared_dist}
        f_{\text{distance}}(dist) = \|dist\|_{2}^2 \tag{$D_\text{squared}$}
    \end{equation*}
\begin{enumerate}
    \item [(3)]An exponential transformation of the distance, as used in \cite{ikkala_breathing_2022}: 
\end{enumerate}
    \begin{equation*} \label{eq:exp_dist}
        f_{\text{distance}}(dist) = \frac{1 - e^{-10\cdot \|dist\|_{2}}}{10}  \tag{$D_\text{exponential}$}
    \end{equation*}

We also compare different \textbf{effort} models.
The first one, denoted as \textit{EJK} in the following, was first presented in \cite{charaja_generating_2024} to simulate realistic arm movements. It consists of three components, motivated by the observation that combining multiple effort terms can improve the plausibility of generated movements~\cite{berret_evidence_2011, wochner_optimality_2020}. The components of \textit{EJK} penalize the mean value of the muscle stimulation commands ($r_\text{energy}$), the jerk, i.e., the change in joint acceleration ($r_\text{jerk}$), and the total work done by the shoulder and elbow ($r_\text{work}$) in terms of angular velocities and torques. 
These components are normalized and weighted by coefficients $c_1, c_2$, and $c_3$, respectively, resulting in the following effort model:
\begin{equation}
    f_{\text{effort}}(r_{\text{energy}}, r_{\text{jerk}}, r_{\text{work}}) = \frac{c_1 r_{\text{energy}} + c_2 r_{\text{jerk}} + c_3 r_{\text{work}}}{c_1 + c_2 + c_3} \tag{EJK} \label{EJK}
\end{equation}

In addition, we consider the three effort models from~\cite{klar_simulating_2023} (\textit{DC}, \textit{CTC}, and \textit{JAC}), where their suitability to predict mid-air pointing movements using a non-RL optimization method (MPC) was examined.
All three models include a penalty for large muscle stimulation commands %
$u$, motivated by the fact that humans seek to minimize their control effort during movement~\cite{todorov_optimal_2002}.
In the following models, these muscle controls are penalized in the squared Euclidean norm (the $r_\text{energy}$ component of the \ref{EJK} model instead penalizes muscle controls using the $l^{1}$-norm~\cite{charaja_generating_2024}).
The \textit{DC} effort model only consists of this penalty term, weighted by a coefficient $c_1$:
\begin{equation}
    f_{\text{effort}}(u) = c_1 \|u\|_{2}^2 \tag{DC} \label{DC}
\end{equation}

The \textit{CTC} model adds a penalty on large changes in commanded torque $\tau$, which is the torque at the joints that directly results from the controlled muscle activations.
This term is motivated by a study from \citeauthor{nakano1999quantitative}~\cite{nakano1999quantitative}, where the minimum commanded torque \textit{change} criterion, i.e., the derivative of $\tau$, was found to be the most effective in explaining the temporal characteristics of actual hand trajectories. %
The \textit{CTC} model is formulated as follows:
\begin{equation}
    f_{\text{effort}}(u, \dot{\tau}) = c_1\|u\|_{2}^2 + c_2 \|\dot{\tau}\|_{2}^2 \tag{CTC} \label{CTC}
\end{equation}

Similarly, the \textit{JAC} model adds a penalty in the kinematic space, namely on the joint accelerations  $x_\text{qacc}$, thus avoiding "jerky" movements. This effort model was introduced in~\cite{nakano1999quantitative} and later found to provide the most comprehensive explanation of mid-air pointing movements~\cite{klar_simulating_2023}. 
In contrast to the \ref{EJK} effort model, this model penalizes the acceleration values themselves instead of their changes (i.e., the jerk). 
The resulting \textit{JAC} model is defined as follows:
\begin{equation}
    f_{\text{effort}}(u, x_{\text{qacc}}) = c_1\|u\|_{2}^2 + c_2\|x_{\text{qacc}}\|_{2}^2 \tag{JAC} \label{JAC}
\end{equation}

\subsection{Tasks}\label{sec:tasks}
\begin{table*}[ht!]
\centering
\begin{tabular}{p{0.12\textwidth}p{0.25\textwidth}p{0.4\textwidth}p{0.15\textwidth}}
\toprule
\textbf{Task} & \textbf{HCI relevance} & \textbf{Challenges} & \textbf{Objective}\\ 
\midrule
Mid-air pointing & Natural direct control method; fundamental component of VR/AR interaction; required for both selection and manipulation tasks & Learn muscle control of shoulder and arm from multi-sensory perception (visuomotor task); keep finger inside target (precise aimed reaching); handle a range of target positions and sizes (multi-goal task) & Finger inside target (dwell time) \\
Object tracking & {Dynamic environments requiring fast visuomotor adaptation} & {Perform continuous error corrections based on perceptual input (closed loop task)} & Finger close to target\\ 
Choice reaction & {Prototype of a physical selection task; sequential button clicking}& {Visuomotor coordination; abstraction; handling contact forces} & Press correct button (force constraints) \\ 
Remote control & {Sequential task; indirect manipulation via joystick control} & {Complex control task of 6th order; requires end-to-end learning of multiple skills (mid-air pointing, joystick control, visuomotor coordination to steer the car position)} & Park car inside target \\ %
Keyboard typing & {Standard bimanual interaction task; sequential task involving physical selection} & {High precision; handling contact forces; large sequences of aimed movements; generalizability across buttons (multi-goal task)} & Press correct buttons (touch) \\ 
\bottomrule
\end{tabular}
\caption{HCI relevance, challenges, and objectives of the five HCI tasks considered in this work.}
\Description{Overview of five HCI tasks: mid-air pointing, object tracking, choice reaction, remote control, and keyboard typing, highlighting their differences in the columns "HCI relevance", "Challenges", and "Objective".}
\label{tab: tasks}
\end{table*}
To develop practical guidelines, which we will propose in Section~\ref{sec:guidelines},
we trained policies for the choice reaction, pointing, and tracking task implemented in the User-in-the-Box (UitB) framework\footnote{\url{https://github.com/User-in-the-Box/user-in-the-box}}.
In each task, we use the default \textit{MoBL Arms Model}~\cite{saul2015benchmarking} with 5 DoFs (three independent shoulder joints, elbow, wrist) and 26 muscles enabled, and provide visual, proprioceptive, and tactile information as input to the agent.
Each episode starts with the arm hanging down (see Figure~\ref{fig:effort_models} (A)).

To validate our guidelines, we define a separate set of movement-based interaction tasks, consisting of a remote control~\cite{ikkala_breathing_2022} and a keyboard typing task~\cite{Hetzel2021}.
In the following, we describe each of the five tasks in more detail.
Table~\ref{tab: tasks} provides an overview of the HCI relevance and objectives for each task, as well as the associated challenges for our RL-based user simulations.

\paragraph{Choice Reaction} In the choice reaction task, the agent faces four colored buttons and a display presenting a stimulus, i.e., one of the four colors. The objective is to press the physical button that corresponds to the displayed color as quickly as possible, with a time limit of four seconds per trial. The agent reaches a goal state when it presses the correct button with sufficient force. This triggers a change of the display color to indicate the start of the next trial. Each episode incorporates 10 trials. Training was conducted for 35 million steps, as further training beyond this threshold did not yield significant improvements. %

\paragraph{Pointing} In the pointing task, the agent must move its index finger to a virtual target sphere of varying radius, positioned in front of the user. A goal state is reached when the agent maintains its fingertip within the target for 500 milliseconds, thereby completing a trial. After either a successful trial or a timeout of four seconds without completion, a new target location is sampled. Each episode contains 10 trials. Each policy has been trained for 50 million steps, after initial tests suggested no further improvements past this.

\paragraph{Tracking} In the tracking task, the agent is required to continuously follow a virtual moving target sphere with its fingertip as precisely as possible. The aim is to point at the target and then keep the fingertip inside the moving sphere. At each timestep, the model is considered to be in a goal state if the fingertip remains inside the target. Each episode lasts ten seconds, regardless of the agent's success. Training was conducted for 50 million steps.

\paragraph{Remote Control} In the remote control driving task, the agent controls acceleration and braking of a toy car by tilting the left joystick of a physical gamepad along a single dimension (forward or backward). The objective is to park the car inside the marked area. 
An obvious goal state is reached when the car is positioned inside this area. Due to the sequential nature of the task, making contact with the joystick can be considered an intermediate goal state. Each episode runs for a fixed duration of ten seconds. Being a more involved task, we heuristically trained each policy for 200 million steps.

\paragraph{Keyboard Typing} For the keyboard typing task, the agent interacts with a physical keyboard, attempting to hit specified keys. Replicating the setup from~\cite{Hetzel2021}, the agent does not use visual perception, but instead obtains the position of the current target key as part of the observation. A goal state is reached when the correct key is successfully pressed, which also terminates the episode. If the agent fails to press the key within three seconds, the episode ends without success. Training was conducted using a single-hand model, after which a mirrored second hand was introduced to enable bimanual typing. Each policy was trained for two million steps.

\subsection{Experimental Design}\label{subsec:exp-design} %

In total, we trained over 500 RL policies across the five tasks introduced in Section~\ref{sec:tasks}, with more than 450 policies dedicated to analyzing the effects of reward function components in the core tasks of pointing, tracking, and choice reaction. 
Throughout, our experimental design sought to balance depth and coverage with the high computational cost of training biomechanical agents. 
We recall that each training may require 12-72 hours to converge.
To ensure robustness and allow for confidence intervals, we initially trained each reward configuration multiple times with seven different %
initial network parameters (weights and biases) for the choice reaction task. 
As preliminary results showed low variance in most settings, we subsequently reduced the number of repetitions, in particular for cases where we observed low performance: three repetitions for policies without task bonus, where we averaged results, and one repetition for policies without a distance term. 
For other tasks, we performed three repetitions except for policies without a distance term, where we limited ourselves to one repetition.

To investigate how individual reward components influence learning outcomes (\ref{item:rq:plausible-movement}), we adopted an incremental strategy: we first trained agents using only the task bonus or distance term, then explored their combinations, and finally introduced the effort model to assess its interaction with the other components. 
Our choice for the respective effort coefficients $c_i$ (see Section~\ref{subsec:reward-components}) was informed by related literature and refined through initial pilot training runs. 
For the sensitivity analysis (\ref{item:rq:sensititivies}), we varied the weight of one reward component while keeping the others fixed, across multiple configurations and tasks. 
As we observed little variance across different formulations of the distance term, our sensitivity analysis only included a single distance term (exponential distance) to avoid redundant computation.
Based on early-stage convergence patterns and similarity in component behavior, we focus on the choice reaction task in our sensitivity analysis, and add observations from the tracking task where differences occurred. 
Finally, to evaluate how our findings generalize beyond the initial task set (\ref{item:rq:generalisabiliy}), we applied our reward design guidelines to the remaining two tasks: keyboard typing and remote control.

To evaluate the trained policies, we ran five episodes per policy. Each episode consisted of 10 trials for the choice reaction and pointing tasks, and 10 seconds of continuous motion for the tracking task. In each episode, the agent starts with the arm hanging down, and targets are randomly sampled, using the same seed for all policies to ensure comparability. We report the mean performance across episodes.

For the choice reaction and pointing tasks, we measured success rate, defined as the percentage of successful trials across the five episodes, and task completion time, which reflects the average time required to complete a trial. 
If a trial is not completed within four seconds, this timeout was recorded as the completion time.
For the tracking task, we evaluated the average distance to target, calculated as the mean Euclidean distance between the fingertip and the surface of the moving target sphere, and the percentage of time the fingertip remained inside the target (measured at a frequency of 20Hz). %

\section{Results}
This section presents the results of training policies with various reward function configurations, in the order of the research questions.
We provide both qualitative %
and quantitative %
evaluations; the former is based on behavioral observations from evaluation videos, while the latter uses task-specific metrics as described in Section~\ref{subsec:exp-design}. 
Parameter values for all experimental conditions (referred to as IDs in the following) are listed in the Supplementary Material, and key outcomes are illustrated in the accompanying video figure. %

\subsection{Plausibility (\ref{item:rq:plausible-movement})}\label{subsubsec:results-qual-plausibility}
We present the results for the choice reaction, pointing, and tracking tasks, in that order.

\subsubsection{Choice Reaction} %
\begin{figure}[!t]
    \centering
    \begin{minipage}[t]{0.49\textwidth}
        \centering
        \includegraphics[width=\linewidth]{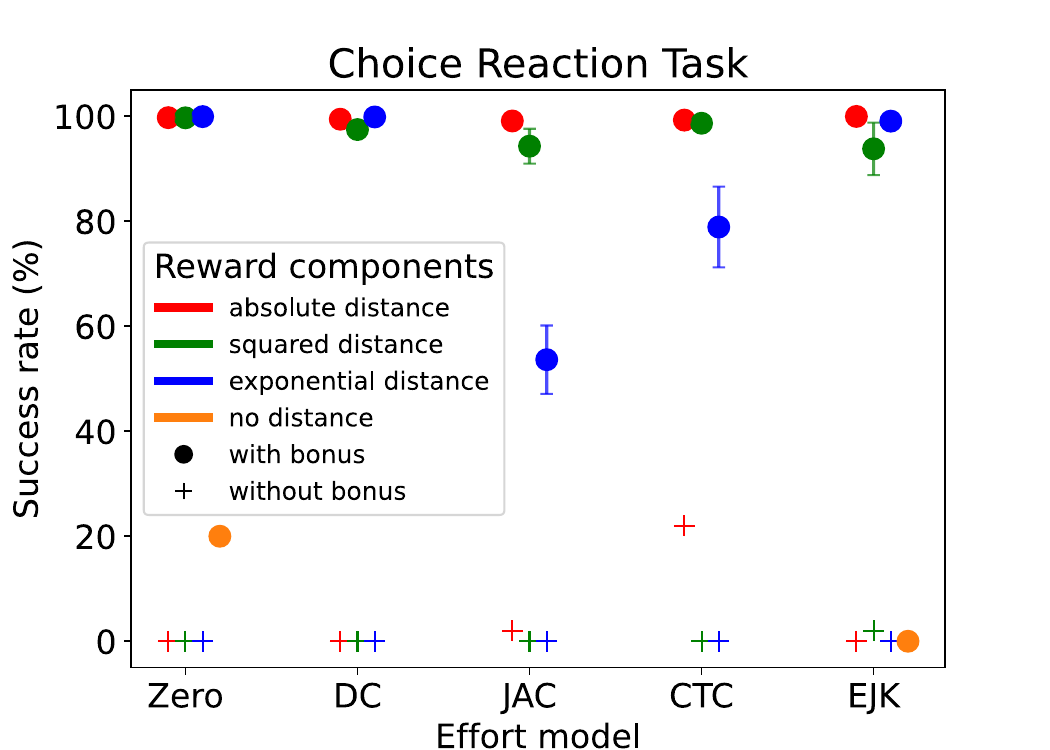}
    \end{minipage}%
    \hspace{0.01em}
    \begin{minipage}[t]{0.49\textwidth}
        \centering
        \includegraphics[width=\linewidth]{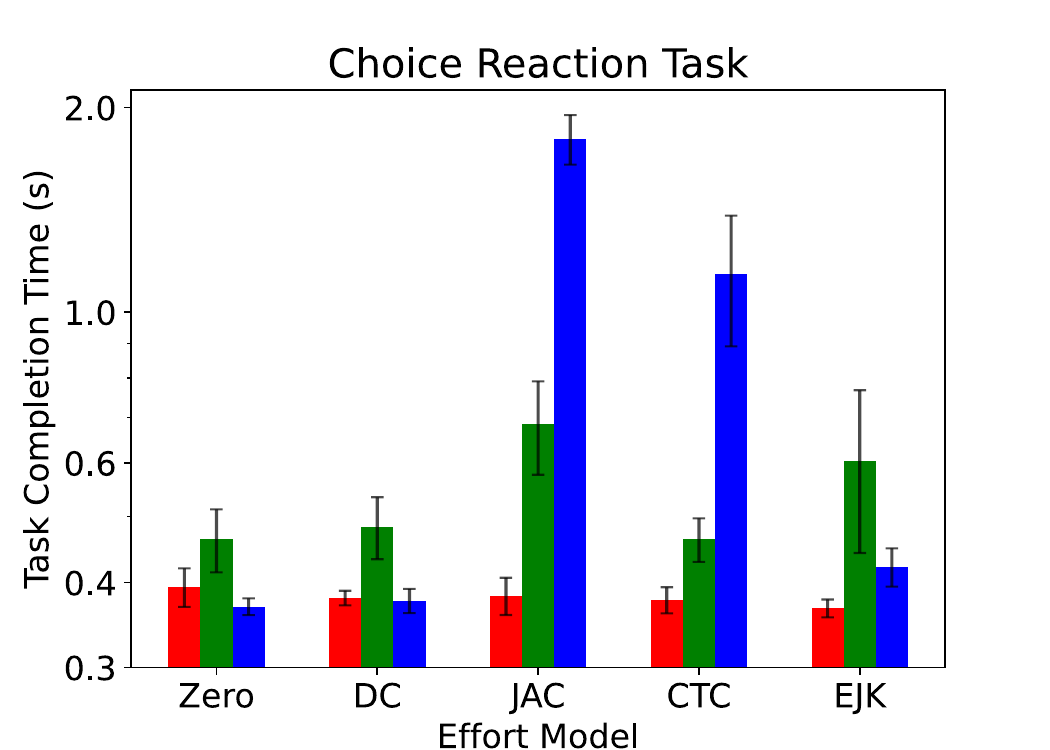}
    \end{minipage}%
    \caption{Success rates (top) and task completion times (bottom) for policies trained with various reward functions in the choice reaction task. %
    Error bars indicate the standard error of the mean.
    Policies trained without either the task bonus (pluses) or distance term (orange circles) achieve success rates below 25\%. Including both terms yields near-perfect success, except for \ref{JAC} and \ref{CTC} effort models. Among high-performing agents, those trained with a squared distance term complete the task considerably slower than with other distance terms.
    }
    \label{fig:success_rate}
    \Description{Success rates (\%) and task completion times (s) for the choice reaction task across reward functions. The top graph shows success rates are highest when both task bonus and distance term are included, with lower success rates under JAC and CTC models. The bottom graph shows task completion times, with agents trained with squared distance term completing tasks more slowly. %
    Error bars indicate the standard error of the mean, with most conditions having relatively small intervals, except for the JAC and CTC models which have larger intervals.}
\end{figure}

\begin{figure*}[!t]
    \centering
    \includegraphics[width=\textwidth]{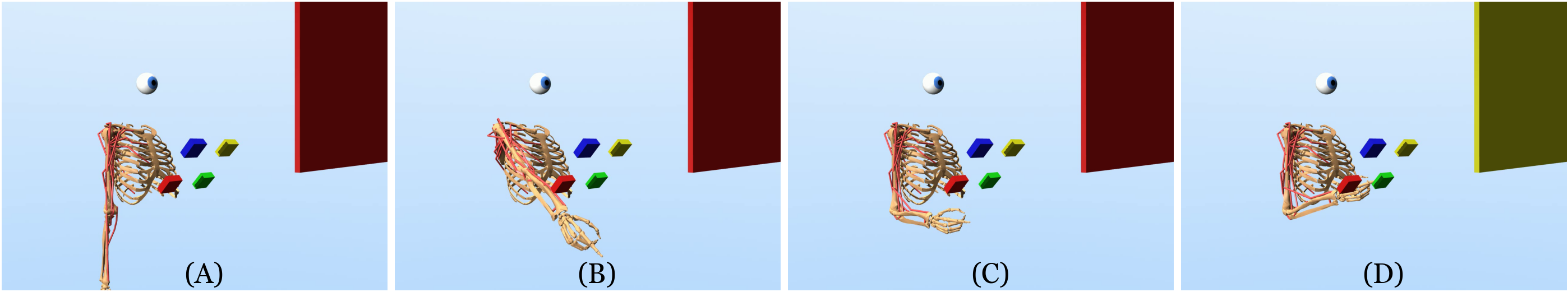}
    \caption{Comparison of (failed) movement patterns of different effort models with (too) high weighting. (A) The \ref{CTC} model does not move at all. (B) The \ref{JAC} model raises the arm but does not bend it. (C) The \ref{DC} model bends the arm but struggles to reach the button. (D) The \ref{EJK} model reaches only the lower buttons.
    }
    \label{fig:effort_models}
    \Description{Movement patterns of a human biomechanical model. The model is positioned in front of four buttons of different colors, with a display representing one of these colors. The agent represents an upper body, a right arm and one eye. A sequence of four images shows the model's arm in different positions: hanging down, stretched out and pointing to a point behind the buttons, bent at a 90-degree angle with hand immediately below the buttons, and finally, with its hand on a lower button.}
\end{figure*}

Figure~\ref{fig:success_rate} shows the success rates and completion times (IDs: 1–35).
With \textbf{task bonus} only (IDs: 31, 32), agents fail to learn generalized behavior -- only the green button is successfully pressed most of the time (orange circle for the "Zero" effort model in Figure~\ref{fig:success_rate} (top)). Movements for the remaining three buttons are noisy and uncoordinated, even when increasing the bonus weight (ID: 32).
With \textbf{distance term} only (pluses for the "Zero" effort model in Figure~\ref{fig:success_rate} (top)), while the agents learn to identify the correct button and move towards it, they fail to press buttons properly and resort to side contacts (IDs: 4–6, 10–12, 16–18, 22–24) as depicted in Figure~\ref{fig:teaser}(B).
\textbf{Effort-only} rewards would lead to no movement, so were excluded.

\textbf{Combining distance and task bonus} results in successful movements (circles for the "Zero" effort model in Figure~\ref{fig:success_rate} (top)) as depicted in Figure~\ref{fig:teaser}(C), with one caveat: the choice of the distance function influences the qualitative behavior for the red button.
The squared and exponential distances (IDs: 25, 27) lead to the agent pressing the red button using the index finger’s proximal phalanx, which often requires multiple attempts. This behavior is not observed with the absolute distance (ID: 26). 

\textbf{Combining distance and effort} (pluses in Figure~\ref{fig:success_rate} (top)), agents mostly fail, either touching (but not pressing) buttons from the side or not reaching them (IDs: 4-6, 10-12, 16-18, 22-24). 
Agents trained with the \ref{JAC} and \ref{CTC} effort models only get close to the button with the absolute distance, whereas the exponential distance "works" for the \ref{DC} and \ref{EJK} models in that they touch the buttons from the side (IDs: 4, 22). 
With increased effort weights, the distinctions between the effort models become more apparent (IDs: 45, 53, 49).
The penalization of joint accelerations in the \ref{JAC} model causes the arm to remain extended, whereas the \ref{DC} model bends the arm and lifts it towards the buttons, but stops below them (Figure~\ref{fig:effort_models} (C)). 

A reward function \textbf{combining only effort and task bonus} (only tested for the \ref{EJK} model; orange circle for \ref{EJK} in Figure~\ref{fig:success_rate} (top)) fails to initiate movement, further emphasizing the need for a complementary, task-specific "guidance" term in the reward function (IDs: 33-35). 
In combination with the task-bonus-only results, we refrained from testing other effort models.

Finally, \textbf{combining all three components} (circles in Figure~\ref{fig:success_rate} (top)), we observe qualitative and quantitative differences between the distance terms. %
Across all effort models, absolute distance leads to the highest success rates ($\geq 96\%$). 
For \ref{JAC} and \ref{CTC} effort models, both absolute and squared distances outperform the exponential distance, with longer completion times for the squared distance (Figure~\ref{fig:success_rate} (bottom)).
However, among the configurations with $\geq 90\%$ success rate, the exponential and absolute distance terms show comparable task completion times. 
Notably, we did not observe a faster task completion for agents trained without an effort model. %
Qualitatively, with the \ref{CTC} effort, the agent is unable to press the upper two buttons and instead keeps the fingertip at one of the lower buttons (IDs: 13-15).
With squared or exponential distance, we observe similar behavior for the \ref{DC} effort model as without effort, where in both cases the red button is pressed with the back of the hand (IDs: 19, 21) -- an effect that vanishes with increasing effort weight.

\subsubsection{Pointing} 

\begin{figure}[!t]
    \centering
    \begin{minipage}[t]{0.49\textwidth}
        \centering
        \includegraphics[width=0.99\linewidth]{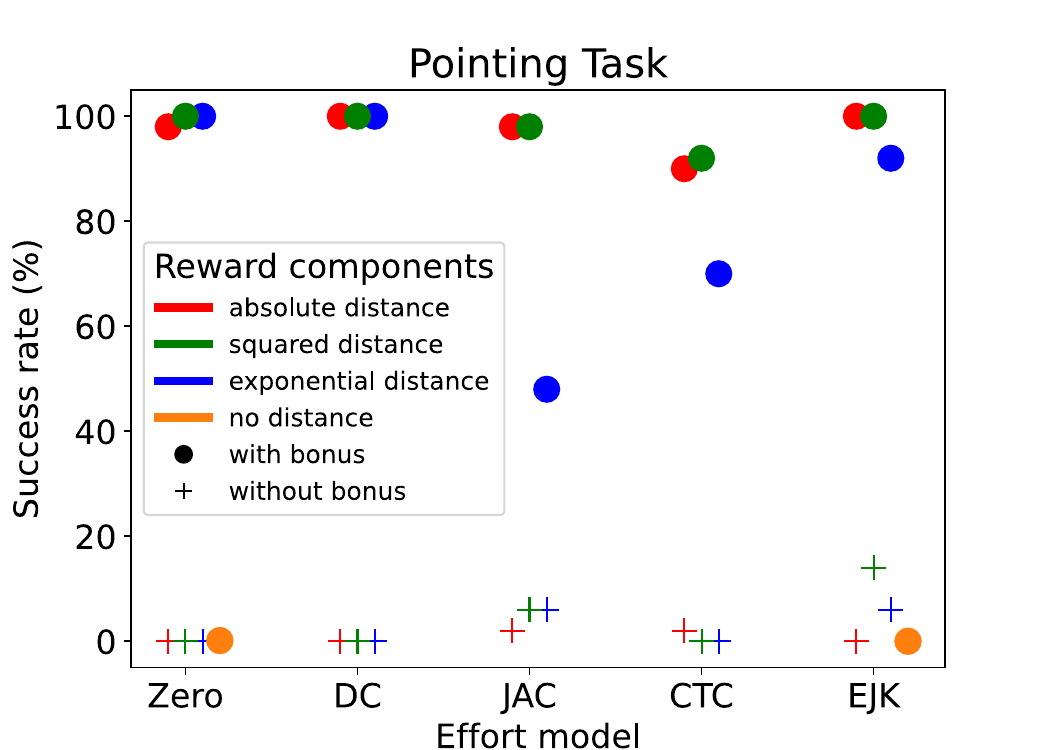}
    \end{minipage}%
    \hspace{0.01em}
    \begin{minipage}[t]{0.49\textwidth}
        \centering
        \includegraphics[width=0.99\linewidth]{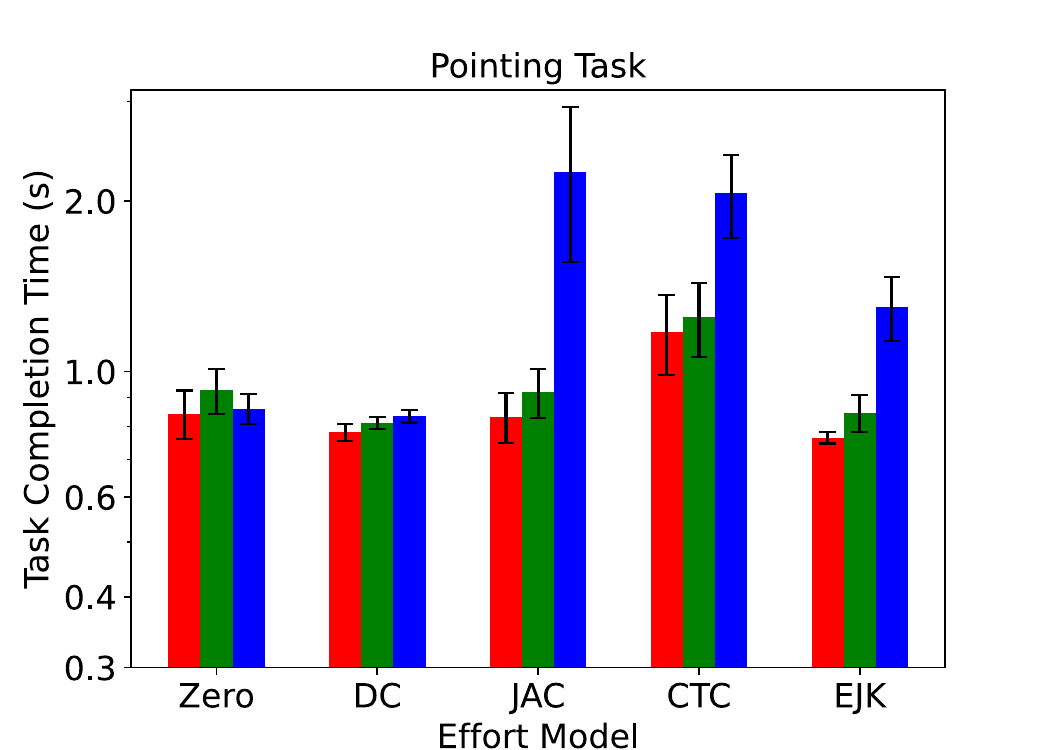}
    \end{minipage}%
    \caption{Success rates (top) and task completion time (bottom) for policies trained with various reward functions for the pointing task. Error bars indicate the standard error of the mean. Policies trained without either the task bonus (pluses) or distance term (orange circles) achieve success rates below 25\%. Including both terms yields near-perfect success, except for \ref{JAC} or \ref{CTC} effort models.}
    \label{fig:success_rates_pointing}
    \Description{Pointing task performance across reward functions. A scatter plot shows success rates (\%) and a bar plot presents task completion time (s). Error bars indicate the standard error of the mean, with JAC combined with exponential distance exhibiting the largest interval. Near-perfect success rates (\%) occur when both task bonus and distance are included, except for JAC and CTC models. Success rates drop below 25\% when either component is missing. Task completion times are longer with squared distance term than with absolute or exponential distance terms among successful policies.}
\end{figure}

Figure~\ref{fig:success_rates_pointing} presents success rates and completion times for agents trained on the \textbf{pointing} task (IDs: 117–150). 
The results are generally similar to those of the choice reaction task.

With \textbf{task bonus} only (orange circle for the "Zero" effort model in Figure~\ref{fig:success_rates_pointing} (top)), agents successfully point at targets close to the arm's initial position, but fail to reach other targets (IDs: 147, 148).
Agents trained with \textbf{distance-only} (pluses for the "Zero" effort model in Figure~\ref{fig:success_rates_pointing} (top)) struggle to stay inside the sphere for the required dwell time, often hovering near the boundary %
(IDs: 144–146).

\textbf{Combining distance and task bonus} (circles for the "Zero" effort model in Figure~\ref{fig:success_rates_pointing} (top)) leads to stable positioning within the sphere ($\geq 90\%$ success rate), with reduced trembling (IDs: 141-143).
In contrast to the choice reaction task, we do not observe any qualitative differences between the distance terms.

\textbf{Combining distance and effort} (pluses in Figure~\ref{fig:success_rates_pointing} (top)) yields agents that exhibit similar behavior as for distance-only. Squared distance combined with any effort model (IDs: 122, 128, 134, 141) and exponential distance combined with the \ref{JAC} or \ref{CTC} effort model (IDs: 126, 132) result in agents that tend to stop short of the target. Other combinations hover over the boundary, not staying inside long enough (IDs: 120, 121, 127, 133, 140, 141).
\textbf{Combining effort and task bonus} (orange circle for the \ref{EJK} effort in Figure~\ref{fig:success_rates_pointing} (top)) results in no movement (IDs: 149, 150).

Finally, \textbf{combining all three components} leads to successful task performance, except for the combination of exponential distance with either \ref{EJK}, \ref{JAC}, or \ref{CTC} effort models, where the agent sometimes keeps the arm outstretched, unable to reach inside the sphere (IDs: 117, 123, 129).
Agents trained without an effort model are again not necessarily faster than those trained with effort model.

\subsubsection{Tracking} 
\begin{figure}[!t]
    \centering
    \begin{minipage}[t]{0.49\textwidth}
        \centering
        \includegraphics[width=\linewidth]{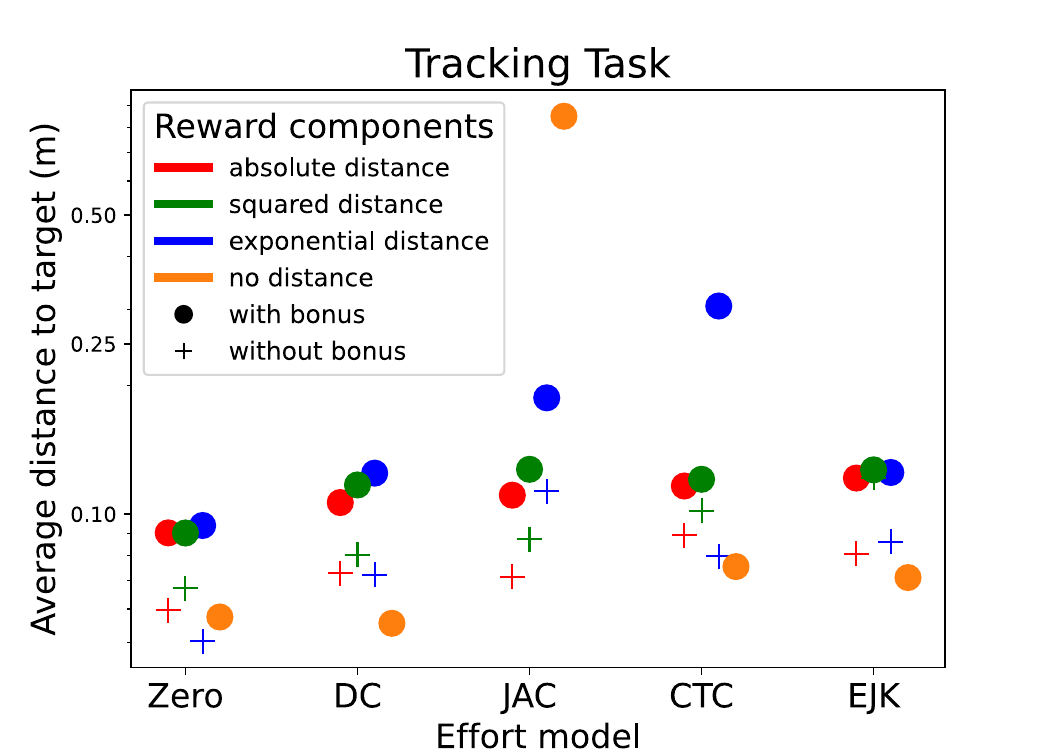}
    \end{minipage}%
    \hspace{0.01em}
    \begin{minipage}[t]{0.49\textwidth}
        \centering
        \includegraphics[width=\linewidth]{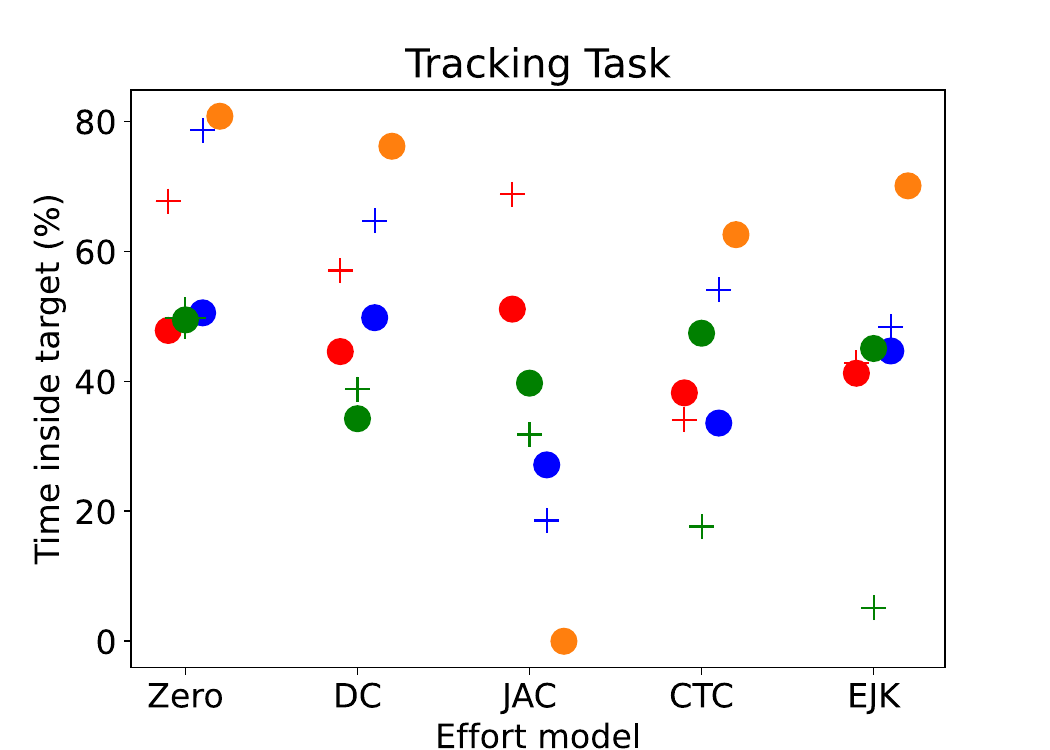}
    \end{minipage}
    \caption{
    Performance comparison of policies trained with various reward functions in the tracking task. The top plot shows the average distance to the target (lower is better), while the bottom plot presents the time spent inside the target area (higher is better). The policy with task bonus and the \ref{JAC} effort model fails to point inside the target at all, similar to the policy combining a task bonus, the \ref{CTC} effort model, and exponential distance.
    }
    \label{fig:tracking_sucess_rates}
    \Description{Two scatter plots compare reward combinations with effort models: Zero, DC, JAC, CTC, and EJK. The top plot shows average distance to the target (m), lower values indicate better performance. JAC with task bonus alone performs worst, followed by combinations of JAC or CTC with task bonus and exponential distance. The closest performance comes from models using distance without task bonus, especially Zero, DC, CTC, and EJK. The bottom plot shows time spent inside the target area as a percentage—higher is better. Again, distance-only combinations for Zero, DC, CTC, and EJK perform best. JAC with task bonus-only never enters the target.}
\end{figure}

Figure~\ref{fig:tracking_sucess_rates} illustrates the average target distance (lower is better; note the logarithmic scale) and the percentage of time spent within the target sphere (higher is better) for the tracking task (IDs: 151–187).

In contrast to the previous tasks, the agent trained with just the \textbf{task bonus} (orange circle for the "Zero" effort model in Figure~\ref{fig:tracking_sucess_rates}) follows the target with high accuracy.
A reward function with only the \textbf{distance term} (pluses for the "Zero" effort model in Figure~\ref{fig:tracking_sucess_rates}) results in agents that stay close to the target but show a lot of trembling (IDs: 181, 182).

Agents trained with \textbf{distance and task bonus} (non-orange circles for the "Zero" effort model in Figure~\ref{fig:tracking_sucess_rates}) remain in close proximity to the target sphere (IDs: 175-177).
However, these agents occasionally lack the precision necessary to maintain their position within the target sphere. 
Notably, \textbf{combining distance and effort models} (pluses in Figure~\ref{fig:tracking_sucess_rates}) results in agents mostly maintaining a closer distance to the target than those with an additional bonus (non-orange circles in Figure~\ref{fig:tracking_sucess_rates}). 
The performance for \textbf{combining effort and task bonus} (orange circles in Figure~\ref{fig:tracking_sucess_rates}) differs between effort models. While agents trained with the \ref{JAC} effort model do not move at all (ID: 187), those trained with the \ref{DC}, \ref{CTC} or \ref{EJK} effort models stay very close to the target sphere (ID: 184-186).

When \textbf{combining all three components}, we observed qualitative differences between the distance terms. Agents trained with the absolute distance tend to remain to the left of the sphere (IDs: 152, 158, 164, 170, 176), while those trained with the squared distance tend to remain to the right (IDs: 153, 159, 165, 171, 177). Agents trained with the exponential distance term exhibit a mixture of these behaviors (IDs: 151, 157, 163, 169, 175).
In combination with task bonus and either the \ref{JAC} or the \ref{CTC} effort model, the exponential distance leads to poorer tracking performance than the other distance terms. %

\subsection{Sensitivity (\ref{item:rq:sensititivies})}

\begin{figure}[!t]
    \begin{minipage}[t]{0.49\textwidth}
        \centering
        \includegraphics[width=\linewidth]{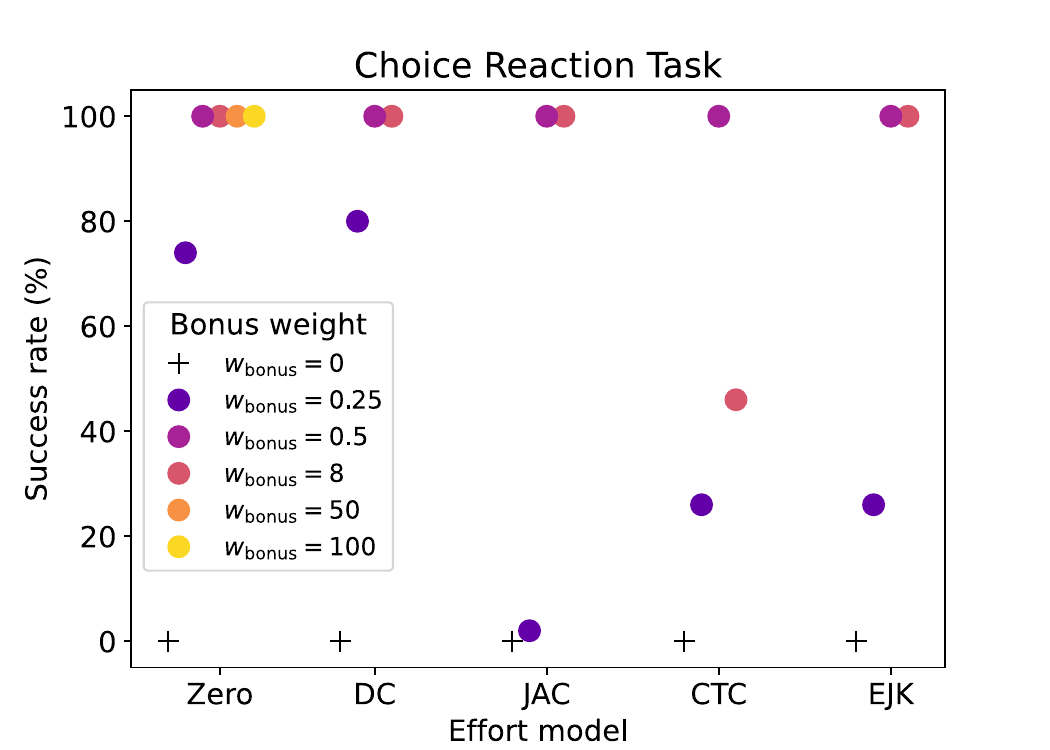}
    \end{minipage}
    \hspace{0.01em}
    \begin{minipage}[t]{0.49\textwidth}
        \centering
        \includegraphics[width=\linewidth]{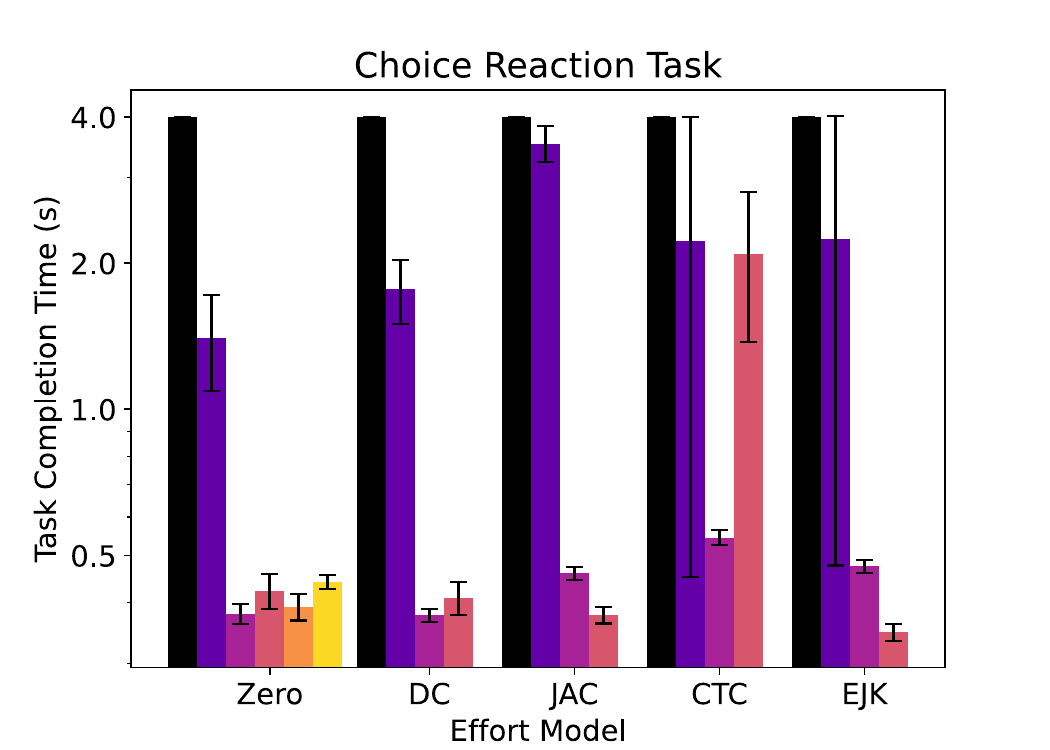}
    \end{minipage}
    \caption{Effect of varying bonus weight on success rate (top) and task completion time (bottom) in the choice reaction task. The task completion time is presented for policies including task bonus and distance. Error bars correspond to the standard error of the mean. Policies trained with a bonus weight greater than or equal to 0.5 achieve success rates of $100\%$, except when using the \ref{CTC} effort model. Lower bonus weights result in lower success rates.}
    \label{fig:choice_reaction_bonus_sensitivities}
    \Description{Two plots compare the effect of bonus weight on the choice reaction task across effort models. The top scatter plot shows success rates, with bonus weights of 0.5 or higher achieving near 100\% success, except for CTC. Bonus weights of 0 or 0.25 result in lower success rates. The lower bar plot shows task completion times, with bonus weights below 0.5 leading to slower completion due to timeouts. Bonus weights above 0.5 result in completion times between 0.2 and 2.6 seconds, except for CTC which takes around 2 seconds. Error bars represent the standard error of the mean, with CTC and EJK each combined with an bonus weight of 0.25 exhibiting larger error bars.}
\end{figure}

\begin{figure}[!t]
     \begin{minipage}[t]{0.49\textwidth}
        \centering
        \includegraphics[width=\linewidth]{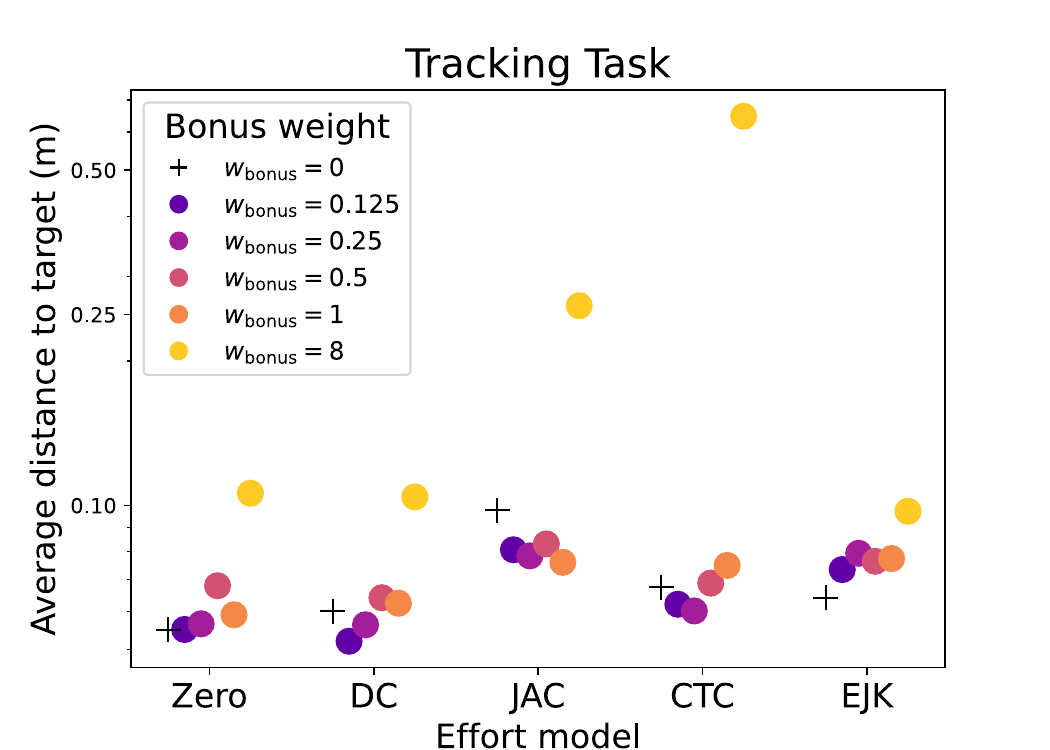}
    \end{minipage}
    \hspace{0.01em}
    \begin{minipage}[t]{0.49\textwidth}
        \centering
        \includegraphics[width=\linewidth]{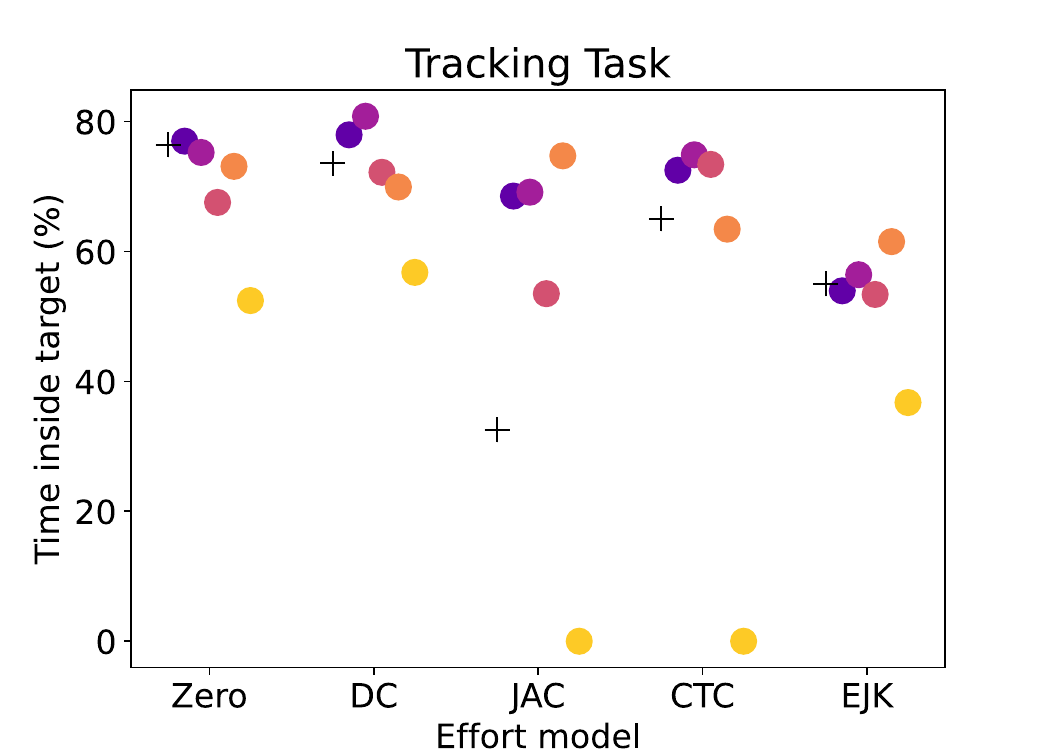}
    \end{minipage}
    \caption{Effect of varying bonus weight on the average distance to the target (top; logarithmic scale) and time spent inside target (bottom) in the tracking task. Small bonus weights ($\leq$ 1) generally result in smaller distances and more time spent inside the target.}
    \label{fig:tracking_success_rates_bonus}
    \Description{Two scatter plots compare the effect of bonus weight on the tracking task. The top plot shows average distance to the target (m), with bonus weights below 8 resulting in distances under 0.1 meters, while a bonus weight of 8 leads to distances up to 0.7 meters, especially for CTC. The lower plot shows time spent inside the target (\%), with bonus weights below 8 resulting in at least 50\% time spent inside, while a bonus weight of 8 leads to 0\% for JAC and CTC, and 35-60\% for other effort models.}
\end{figure}

We analyze the sensitivity of each reward component %
by varying one weight while keeping the others fixed. 
Based on the results from Section~\ref{subsubsec:results-qual-plausibility}, we decided to focus on exponential distance as it showed the greatest variability between effort models, but otherwise performed similarly to absolute and squared distance.
For the choice reaction task, we use $w_\text{distance} = 1$ (exponential), $w_\text{effort} = 1$, and $w_\text{bonus} = 8$ as our baseline. 
For the tracking task, we select $w_\text{bonus} = 0.5$ as baseline, as it constitutes a promising candidate for high success rates based on initial findings.

\subsubsection{Bonus Weight} 
Figure~\ref{fig:choice_reaction_bonus_sensitivities} illustrates the success rates and completion times for the \textbf{choice reaction} task (IDs: 1-30, 66–84), %
for varying bonus weights $w_{\text{bonus}} \in \{0, 0.25, 0.5, 8, 50, 100\}$. 
Agents trained with a bonus weight of 0 or 0.25 exhibit low success rates, typically failing to press buttons despite approaching them (IDs: 71–76, 78, 80, 82, 84). %
In contrast, a bonus weight of 0.5 or 8 (IDs: 1–3, 7–9, 14–15, 19–21, 25–28, 68-70, 77, 79, 81, 83)
yields high success (and low task completion time) across all effort models except \ref{CTC}: for 0.5, the agent first touches the buttons from the side before succeeding (ID: 77); for a value of 8, the success rate diminishes.
While no negative effect of a large bonus on success rate was observed for policies trained without effort term, too large bonus weights can slow down learning. For instance, weights of 8, 50, and 100 lead to convergence in 12M, 17M, and 24M steps, respectively (IDs: 31, 66, 67).

Figure~\ref{fig:tracking_success_rates_bonus} depicts the success rates and completion times for the \textbf{tracking} task (IDs: 151-180, 188-207), %
for varying bonus weights $w_{\text{bonus}} \in \{0, 0.125, 0.25, 0.5, 1, 8\}$. %
In contrast to the choice reaction task, a "large" bonus weight (in this case $8$) does degrade performance compared to lower weights, including 0 (note the logarithmic scale for the average distance). 
The \ref{JAC} and \ref{CTC} models seem to profit most from a carefully chosen bonus weight.

\subsubsection{Distance Weight}
\begin{figure}[!t]
    \begin{minipage}[t]{0.49\textwidth}
        \centering
        \includegraphics[width=\linewidth]{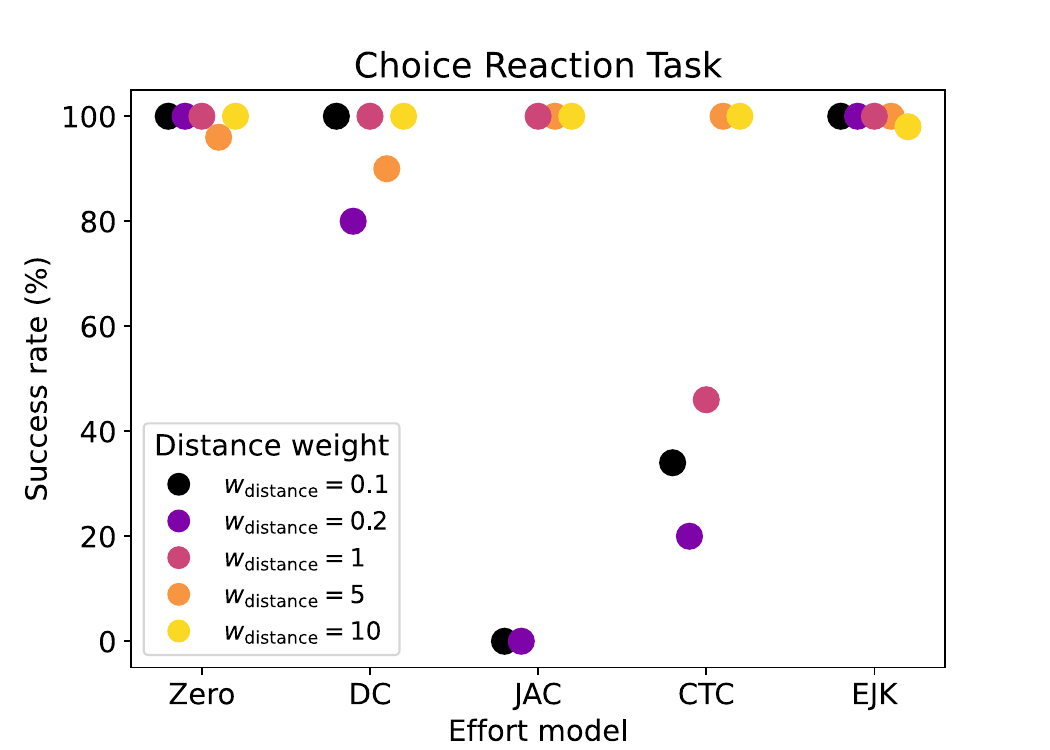}
    \end{minipage}
    \hspace{0.01em}
    \begin{minipage}[t]{0.49\textwidth}
        \centering
        \includegraphics[width=\linewidth]{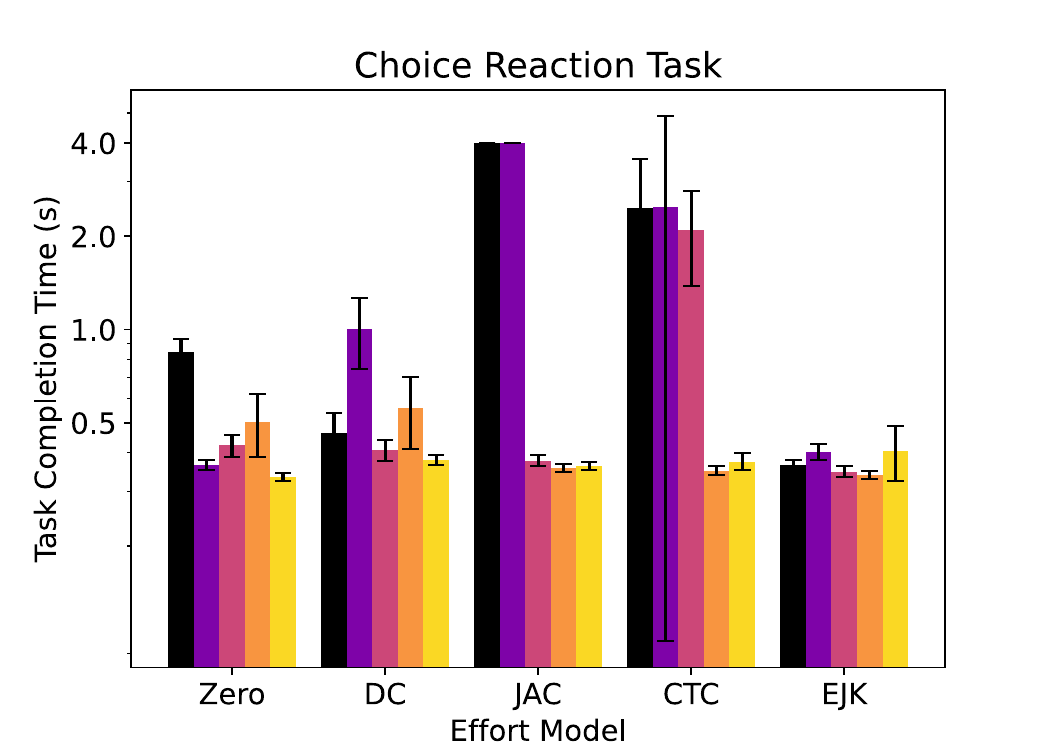}
    \end{minipage}
    \caption{Effect of varying distance weights on success rate (top) and task completion time (bottom) in the choice reaction task. Error bars represent the standard error of the mean. Combined with the \ref{CTC} or \ref{JAC} effort models, small distance weights ($\leq 1$ and $\leq 0.2$, respectively) may lead to failure.}
    \label{fig:choice_reaction_dist_sensitivities}
    \Description{Two plots compare the effect of distance weight on the choice reaction task across effort models. The top scatter plot shows success rates (\%), with distance weights below 1 resulting in low success rates for JAC (0\%) and CTC (20-40\%), while other combinations achieve over 80\% success, except CTC with distance weight 1 (50\%). The lower plot shows task completion times (s), with CTC and JAC having the longest times (2-4 seconds) for distance weights below 1, and other times varying between 0.2 and 1 second, with no clear pattern. Error bars represent the standard error of the mean, with CTC combined with small distance weights resulting in the largest interval.}
\end{figure}

\begin{figure}[!t]
     \begin{minipage}[t]{0.49\textwidth}
        \centering
        \includegraphics[width=\linewidth]{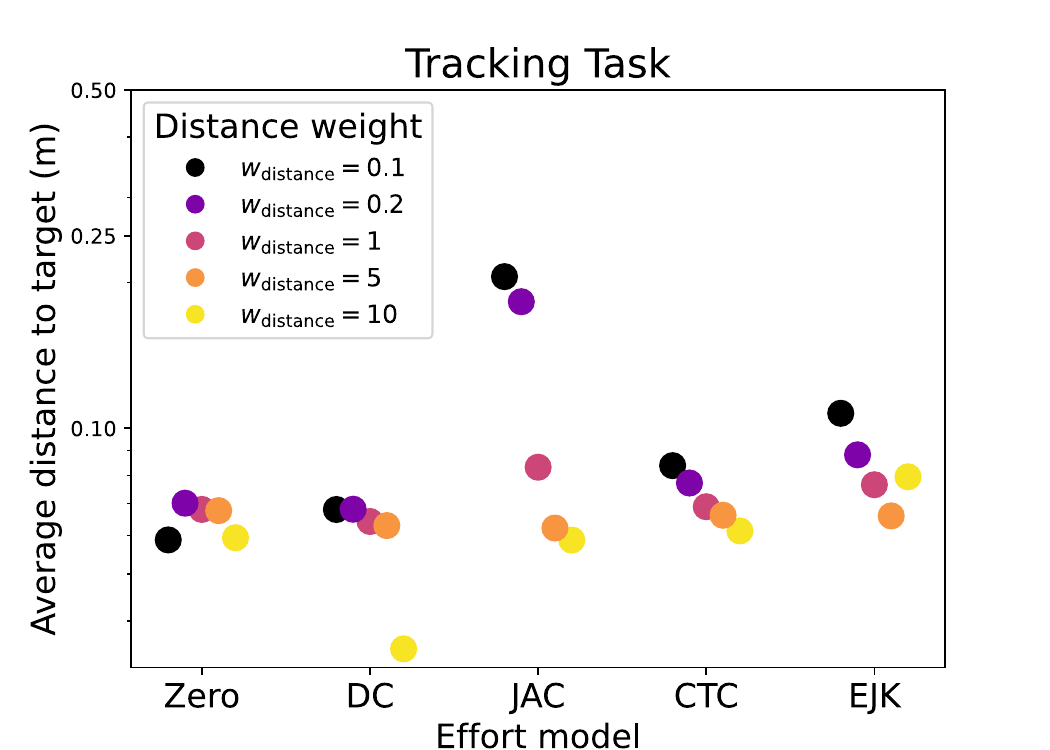}
    \end{minipage}
    \hspace{0.01em}
    \begin{minipage}[t]{0.49\textwidth}
        \centering
        \includegraphics[width=\linewidth]{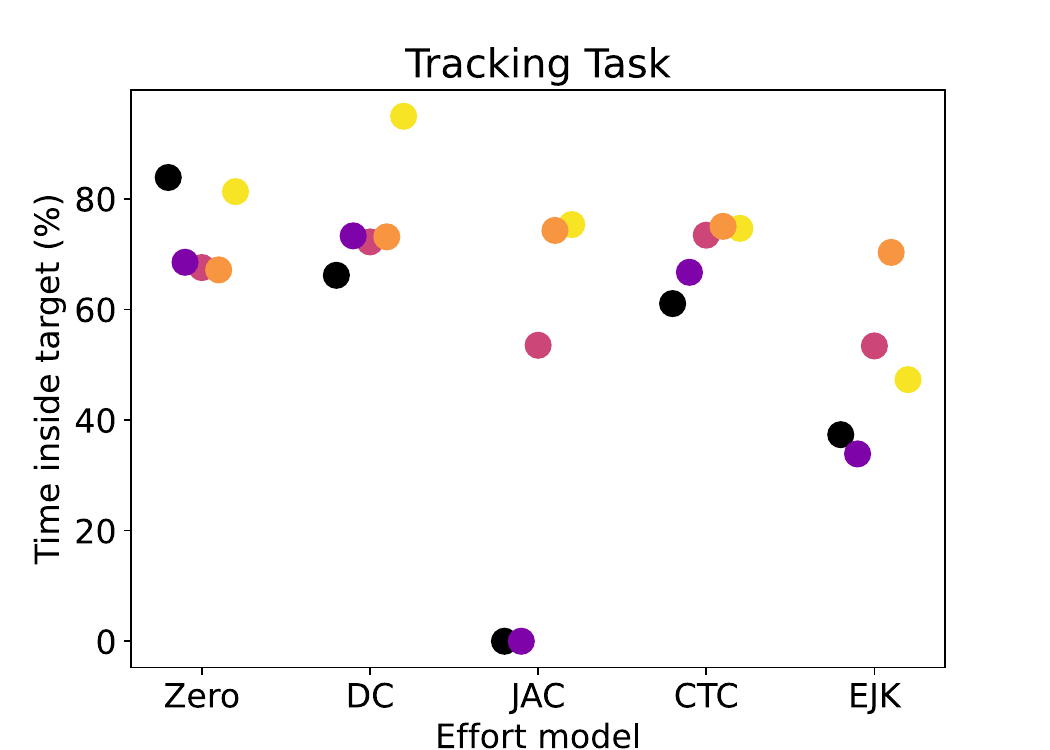}
    \end{minipage}
    \caption{Effect of varying distance weights on the average distance to the target (top; logarithmic scale) and time spent inside target (bottom) in the tracking task. Combined with the \ref{JAC} effort model, small distance weights ($\leq 0.2$) lead to agents not being able to reach the moving target. Overall, small distance weights lead to a less time inside the target.}
    \label{fig:tracking_success_rates_distance}
    \Description{Two scatter plots compare the effect of distance weight on the tracking task across effort models. The top plot shows average distance to the target (m), with JAC and distance weights below 1 resulting in the largest distances (over 0.15 meters), while DC with a distance weight of 10 achieves the smallest distance (near zero). Other configurations result in distances between 0.06 and 0.12 meters. The lower plot shows time spent inside the target (\%), with JAC and distance weights below 1 resulting in 0\% time inside, while DC with a distance weight of 10 achieves over 85\%. Other agents exhibit times between 30 and 85\%.
    }
\end{figure}

Figure~\ref{fig:choice_reaction_dist_sensitivities} shows the effects of varying the distance weight ($w_{\text{distance}} \in \{0.1, 0.2, 1, 5, 10\}$) in the \textbf{choice reaction} task (IDs: 36–65). 
We recall that agents are trained with the exponential distance term and a fixed bonus weight of $w_{\text{bonus}} = 8$.
A distance weight larger than $1$ has limited impact on policies (IDs: 42–48, 50, 51, 54, 55, 58, 59, 62, 63).
Lower distance weights lead to lower success for some effort models (outstretched arm for \ref{JAC}, bent arm for \ref{CTC}), and may lead to larger completion times (zero effort), where agents tap the button with the back of the hand, requiring several attempts to press a button (IDs: 45–47).
The \ref{EJK} effort model is the most robust to changes in the distance weight\new{, while the \ref{CTC} effort model combined with a distance weight of 0.2 exhibits a notably high standard error of the mean.}

For the \textbf{tracking} task (Figure~\ref{fig:tracking_success_rates_distance}, IDs: 208–227), results indicate that increasing the distance weight (slightly) improves accuracy (note the logarithmic scale) and time inside target for agents trained with any non-zero effort models, particularly for the \ref{JAC}, \ref{CTC}, and \ref{EJK} effort models. 
For the lowest distance weights, the \ref{JAC} model tries to reach the target with an outstretched arm, while the \ref{CTC} model follows the target shakily.

\subsubsection{Effort Weight}

\begin{figure}[!t]
    \begin{minipage}[t]{0.49\textwidth}
        \centering
        \includegraphics[width=\linewidth]{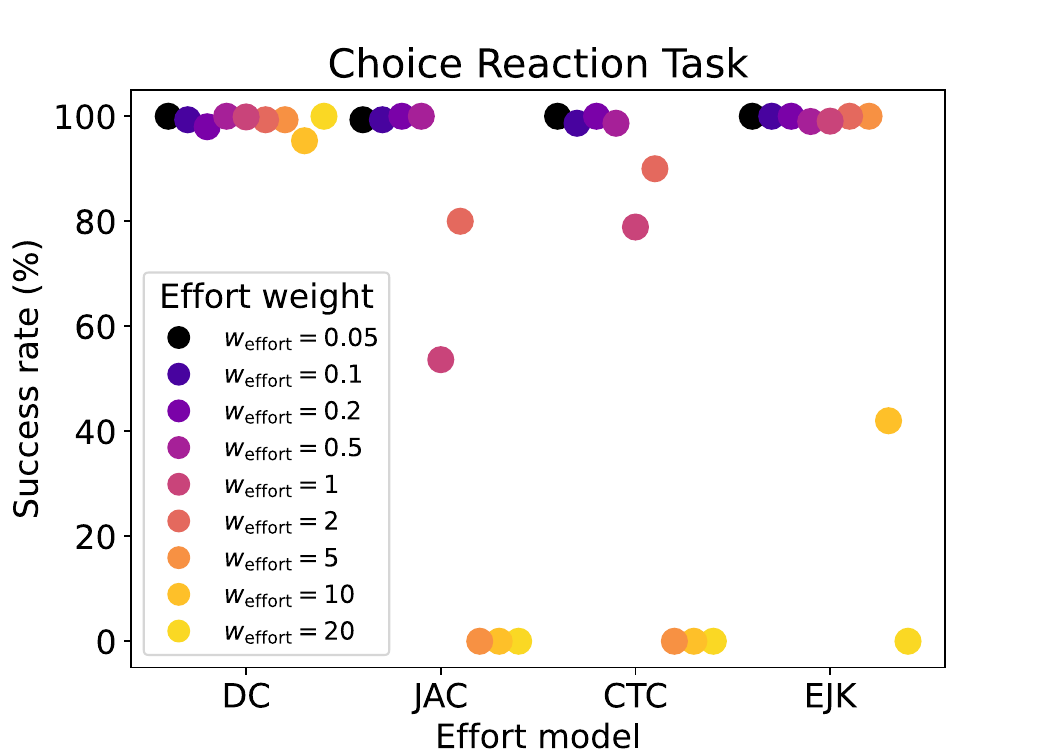}
    \end{minipage}
    \hspace{0.01em}
    \begin{minipage}[t]{0.49\textwidth}
        \centering
        \includegraphics[width=\linewidth]{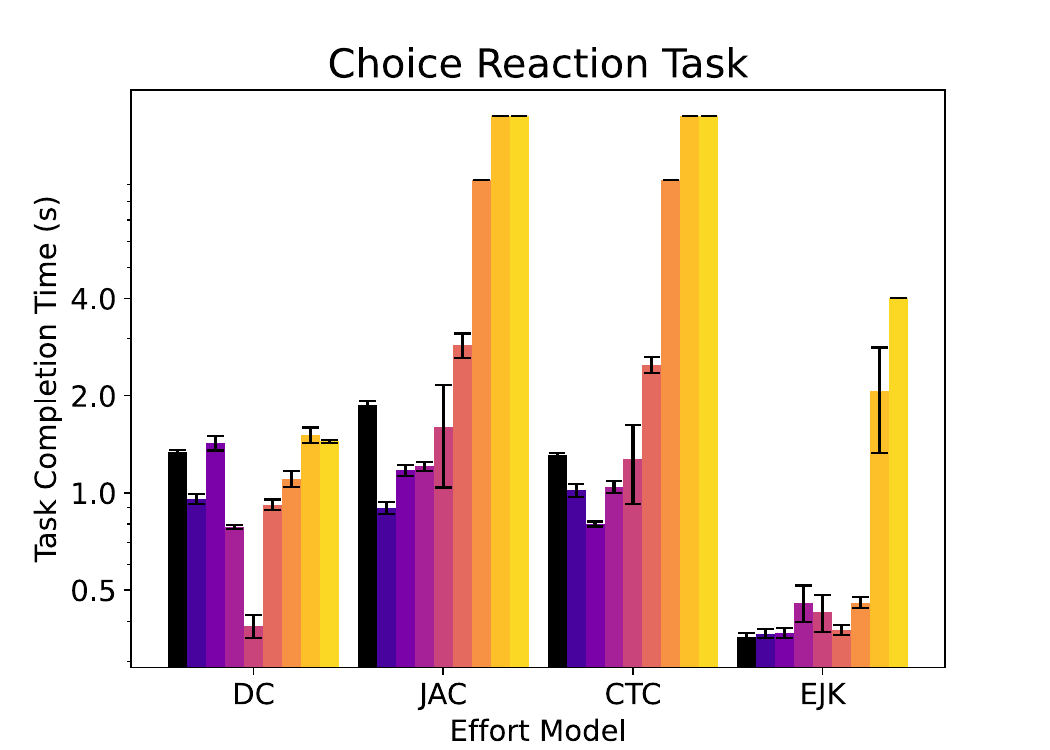}
    \end{minipage}
    \caption{Effect of varying effort weight on success rate (top) and task completion time (bottom) in the choice reaction task. Error bars represent the standard error of the mean. Larger effort weights lead to task failure for the \ref{JAC}, \ref{CTC}, and \ref{EJK} effort models. %
    }
    \label{fig:choice_reaction_effort_weights}
    \Description{Two plots compare the effect of effort weight on the choice reaction task across effort models. The top scatter plot shows success rates, with effort weights above 2 resulting in 0\% success for JAC and CTC, and also for EJK with a weight of 20. Effort weights of 1-2 for JAC and CTC, and 10 for EJK, achieve success rates between 40-90\%. Other agents achieve near 100\% success. The lower bar plot shows task completion times, with JAC, CTC, and EJK models having times over 2 seconds with effort weights above 5, while others have times between 0.3-2 seconds, with no clear pattern. Error bars represent the standard error of the mean, with JAC combined with an effort weight of 1 and EJK combined with an effort weight of 10 exhibiting the largest intervals.
    }
\end{figure}

Figure~\ref{fig:choice_reaction_effort_weights} shows the impact of varying effort weights ($w_\text{effort} \in 0.05, 0.1, 0.2, 0.5, 1, 2, 5, 10, 20$) on success rate and task completion time in the \textbf{choice reaction} task (IDs: 85–116). 
We recall that $w_{\text{distance}} = 1$ and $w_{\text{bonus}} = 8$. %
Too large effort weights ($\geq 2$, depending on the effort model) tend to reduce success rates and negatively affect completion time. %
Effort weights greater than $10$ prevent the agent with the \ref{JAC}, \ref{CTC}, and \ref{EJK} models from completing the task (IDs: 85–108), or allow them to press only the button closest to their initial position (ID: 36).
Lowering effort weights does not necessarily lead to lower completion times, and weights below $0.2$ may introduce a tendency to stick to sub-optimal movements and behaviors, e.g. with one agent occasionally struggling to correctly press the red button (ID: 43)
The \ref{DC} models seems most robust to effort weight changes.

\begin{figure}[!t]
     \begin{minipage}[t]{0.49\textwidth}
        \centering
        \includegraphics[width=\linewidth]{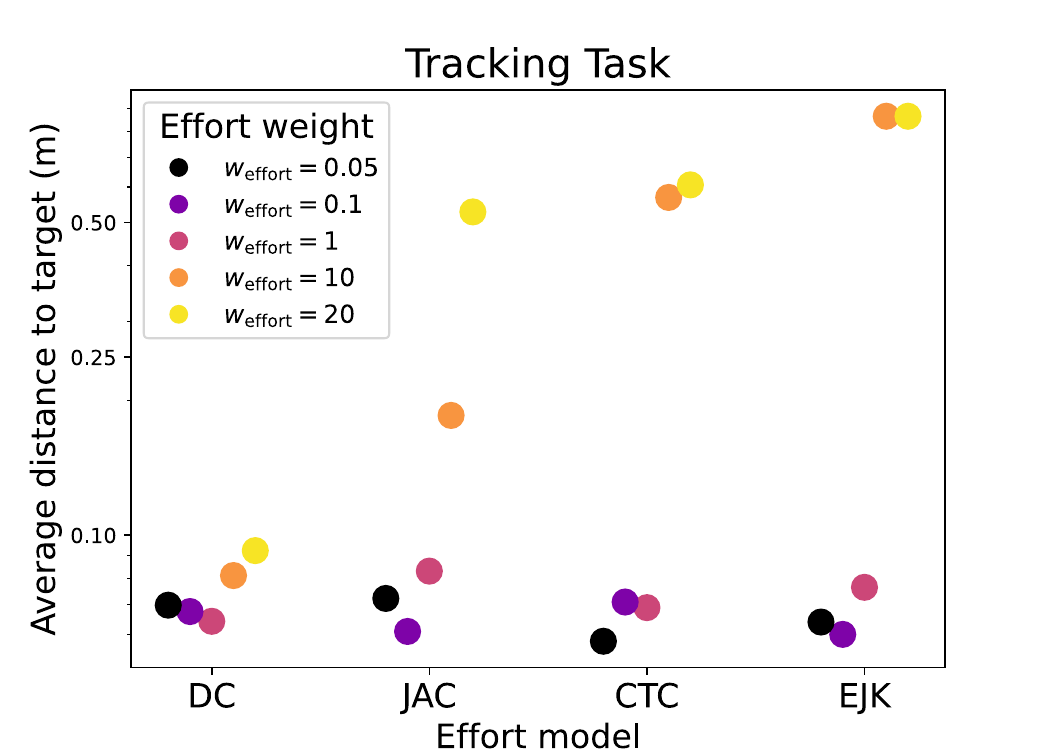}
    \end{minipage}
    \hspace{0.01em}
    \begin{minipage}[t]{0.49\textwidth}
        \centering
        \includegraphics[width=\linewidth]{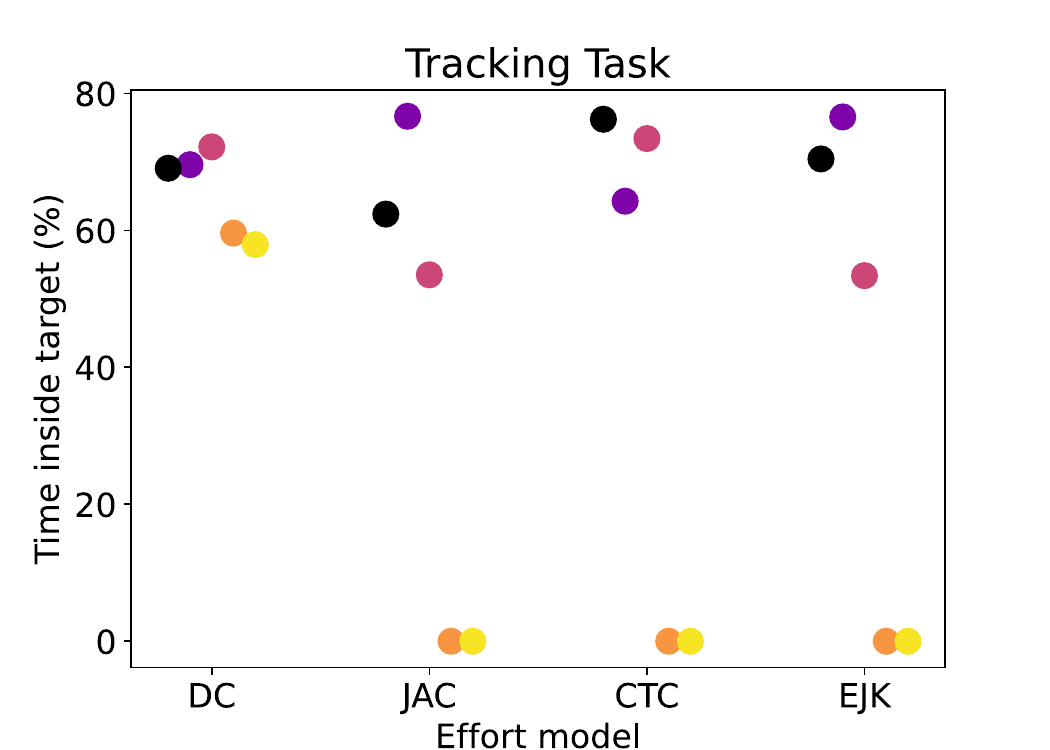}
    \end{minipage}
    \caption{Effect of varying effort weight on the average distance to the target (top; logarithmic scale) and time spent inside target (bottom) in the tracking task. Large effort weights for the \ref{JAC}, \ref{CTC}, and \ref{EJK} effort models lead to agents that fail to point inside the target at all.}
    \label{fig:tracking_effort_sens}
    \Description{
    Two scatter plots compare the effect of varying effort weights (0.05, 0.1, 1, 10, 20) for the effort models DC, JAC, CTC and EJK on the average distance to the target (meters) and time spent inside the target (seconds) in the tracking task. The CTC or EJK effort models combined with effort weights larger than or equal to 10 and the JAC model with a distance weight of 20 result in average distances greater than 0.5 meters, The JAC effort model with an effort weight of 10 leads to 0.23 meters and the others achieve average distances smaller than 0.1 meters. The lower plot shows the percentage of time spent inside the target, effort weights larger than 10 combined with the JAC, CTC or EJK effort model lead to 0\% time spent inside the target. The others spend at least 50\% of the time inside the target.
    }   
\end{figure}

Similarly, in the \textbf{tracking} task (Figure~\ref{fig:tracking_effort_sens}, IDs: 228–259), large effort weights lead to increased failure to point inside the target (again note the logarithmic scale). %
Effort weights greater than $10$ show similar behavior to distance weights that are too low; in the \ref{EJK} model, too large effort weights cause the agent to not move at all. 
With small effort weights, agents tend to stay closer to the target. %
As with the choice reaction task, the \ref{DC} model appears to be least sensitive to changes in the effort weight.

\subsection{%
Guidelines}\label{sec:guidelines}

\begin{figure*}[!t]
    \centering
    \includegraphics[width=\textwidth]{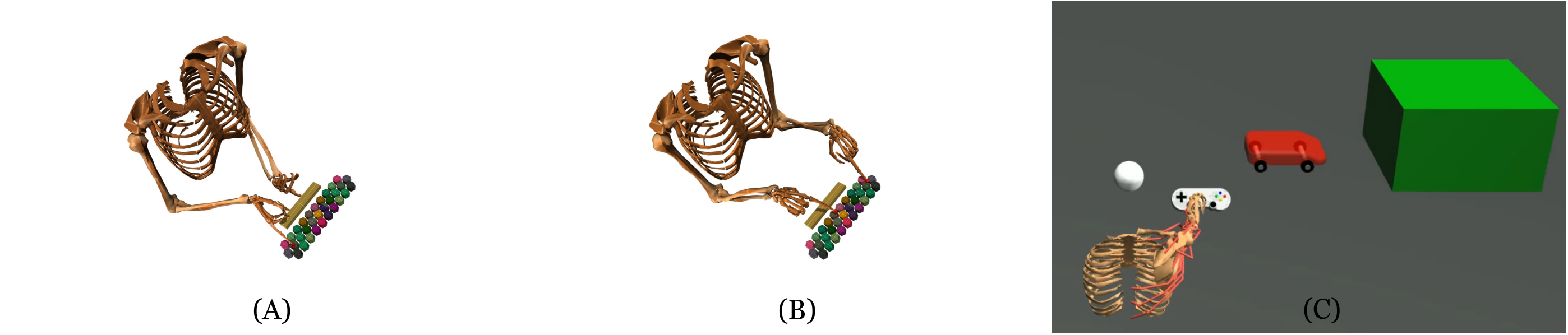}
    \caption{Illustrations of the keyboard typing task: (A) An agent trained with a distance-only reward attempting to press the keys from below, (B) an agent successfully pressing the keys, and (C) the remote control task.}
    \label{fig:keyboard_remote_img}
    \Description{The figure is divided into three illustrations. The leftmost two images illustrate different movement patterns of a human biomechanical model in a keyboard typing task. The model consist of an upper body and two arms and is in front of a virtual keyboard. In the leftmost image the model places its left hand on the lower side of the keyboard and its right hand on the upper side. The middle image presents the model touching the keyboard with both fingers from above. In the rightmost image the agent represents an upper body, a right arm and one eye and outstretches the arm to touch the joystick of a gamepad. In front and right of the agent there is a toy car which is again in front of a target box.
    }
\end{figure*}

Based on the above results, we distill guidelines, \new{meant to be followed sequentially}, to assist in the design of a reward function. 
The guidelines are presented in the order in which we suggest they should be followed.

First, we find that a guiding term, such as a distance-based reward, is crucial for training agents to perform goal-directed HCI tasks. Without this component, models in the choice reaction and pointing tasks fail to approach the target and instead perform random uncontrolled movements. Incorporating a distance term allows agents to effectively move towards the target.
We therefore propose the following guideline: %
\begin{enumerate}
    \item[\textbf{G1}] Integrate guidance components, such as distance-based rewards, to effectively guide the agent toward task completion. 
\end{enumerate}

Second, our findings highlight the need for a task completion bonus to achieve high success rates.
Without this bonus, agents often adopt suboptimal strategies, such as touching buttons from the side (choice reaction), or exhibit excessive trembling, which prevents them from keeping the fingertip inside the target sphere for the required dwell time (pointing).
We thus recommend:
\begin{enumerate}
   \item[\textbf{G2}] Include a task bonus to ensure task success. %
\end{enumerate}

The value of the bonus weights is also crucial. 
Setting the bonus weights in the choice reaction task too small results in agents that move the finger close the button but fail to successfully press them. %
While increasing the bonus weight eventually enables proper task execution, too large bonus weights can slow down the training. 
When the task bonus can be awarded multiple times per trial, as it was the case for tracking, we observed a preference toward smaller bonus weights. In this case, a too large bonus weight can lead to inaccuracies.
\begin{enumerate}%
    \item[\textbf{G3}] Adapt the bonus weight:
    \begin{itemize}
    \item If the trial terminates when the bonus is given: increase the bonus weight if the agent reaches the target but fails to complete the trial successfully.
    Reduce the bonus weight if the agent requires excessive training steps.
    \item If the bonus is awarded multiple times per trial, in addition to a guidance component: keep the bonus weight low and adjust for accuracy. %
    \end{itemize}
\end{enumerate}

The results also suggest that effort models are not always necessary to generate successful movement trajectories. While we observed some irregularities in the choice reaction task (trying to repeatedly press the button with the back of the hand), none occurred in the pointing and tracking tasks. 
Hence, 
we suggest:
\begin{enumerate}%
    \item[\textbf{G4}] Try without effort models first. If instabilities arise (e.g., excessive trembling), add an effort model. %
\end{enumerate}

Effort-based penalties offer a mechanism to encourage smoother or more efficient movement. However, they must be balanced carefully with distance-based rewards. A sufficiently high effort weight can suppress undesirable behavior like erratic motion (e.g., pressing with the back of the hand), but overly strong penalties may discourage movement altogether. Striking the right balance is task dependent and typically requires iterative weight tuning based on first results.
We therefore recommend:
\begin{enumerate}%
    \item[\textbf{G5}] Dynamically balance distance and effort weights: reduce the effort weight or increase the distance weight if the agent fails to approach the target; do the opposite if movement instability is observed.
\end{enumerate}

\subsection{Case Studies: Keyboard Typing and Remote Control Driving}

We evaluate our proposed reward design guidelines on two HCI tasks: keyboard typing \cite{Hetzel2021} and remote control \cite{ikkala_breathing_2022}, as shown in Figure~\ref{fig:keyboard_remote_img}. 
Unlike prior work that relies on curriculum learning and multi-stage training processes \cite{Hetzel2021}%
, our goal is to demonstrate that an appropriately designed reward function, utilizing our guidelines, can lead to successful policy learning with a straightforward %
training setup. \new{Notably, the keyboard typing task is implemented and evaluated outside the user-in-the-box framework, demonstrating the broader applicability of our approach.}
We train agents for two million steps on the keyboard typing task and 200 million steps on the remote control task.

\begin{figure}[!t]
    \begin{minipage}[t]{0.5\textwidth}
        \centering
        \includegraphics[width=0.85\linewidth]{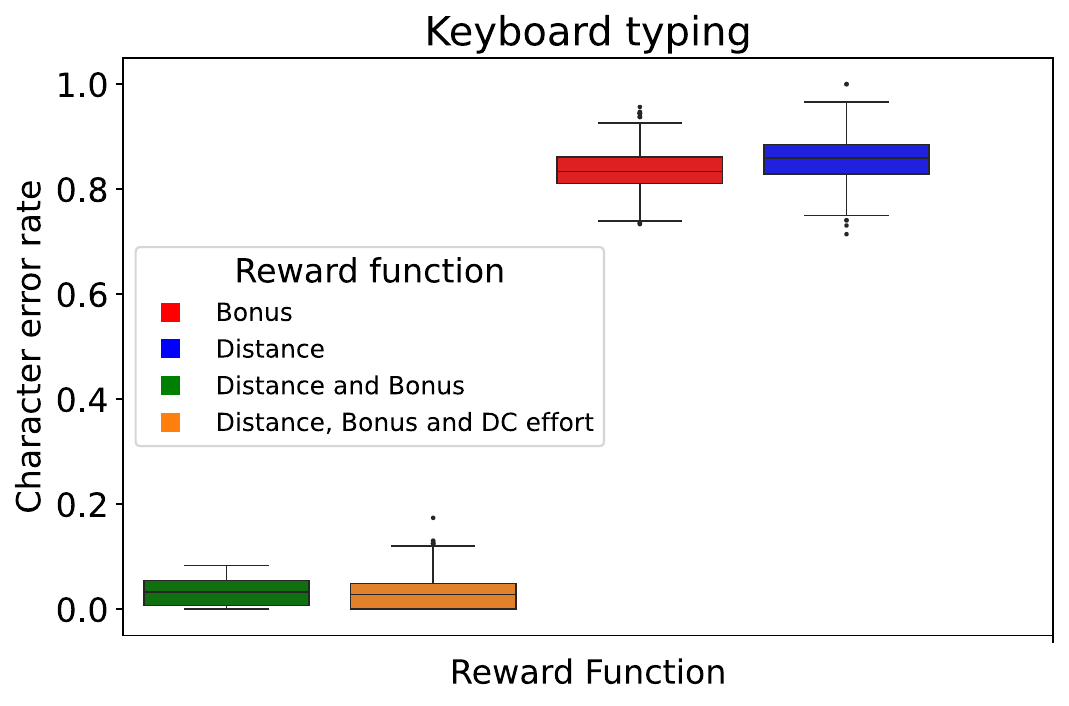}
    \end{minipage}
    \caption{Comparison of character error rates for policies trained with various reward functions in the keyboard typing task (lower is better).}
    \label{fig:Keyboard_Typing_success_rates}
    \Description{The boxplot shows character error rates for four reward functions: bonus, distance, distance and bonus, and distance, bonus, and DC effort. The reward function with distance and bonus exhibits a character error rate between 0 and 0.1. The agent trained with distance, bonus and the DC effort model exhibits similar values with outliers up to a character error rate of 0.2. Distance-only exhibits the highest values between 0.7 and 1 and bonus-only achieves similar values between 0.7 and 0.95.}
\end{figure}

\paragraph{Keyboard Typing} For the \textbf{keyboard typing} task, performance is measured using the character error rate over 500 stimulus phrases (each fewer than 40 characters long), which are sampled from the same dataset as used in~\cite{Hetzel2021}; results are shown in Figure~\ref{fig:Keyboard_Typing_success_rates}.

Following our guidelines, the initial reward function combined a distance term penalizing the distance between the fingertip and target key (G1) with a task bonus for correctly pressing the key (G2). Since the bonus immediately terminates the episode, we chose a large bonus weight $w_\text{bonus} = 30$ (G3).
The resulting agent successfully learns the task and achieves character error rates $\leq 10\%$.
\new{Importantly, prior work required a complex two-stage curriculum approach and 5 million steps to reach a character error rate below 10\% \cite{Hetzel2021}. With our reward design, we achieve comparable performance with a character error rate under 10\% in only 2 million steps without specialized training procedures, demonstrating the effectiveness of our guidelines.}

Recognizing that effort models can be beneficial for shaping specific behavior, we also evaluate agents including the \ref{DC} effort in the reward function, previously shown to perform robustly with respect to effort weights in related tasks. Using the same effort weight as in our earlier experiments ($w_\text{effort} = 1$), the resulting agents again succeed in the typing task, achieving character error rates below 20\%, confirming that our guidelines generalize well to this setting.%

To further validate the necessity of the distance and bonus reward components, we conduct more studies. The agent trained with only the task bonus (no distance guidance) fails to learn the right behavior, producing arbitrary movements and a character error rate above 75\%. Conversely, the agent trained with only the distance term approaches the target keys but fails to press them, sometimes trying to press them from below, see Figure~\ref{fig:keyboard_remote_img} (A). This policy results in a character error rate of over 75\%. These findings provide empirical support for the necessity of both guidance and task completion incentives, as outlined in guidelines G1 and G2, with a bonus weight not too large to slow down the learning process (G3).

\begin{figure}[!t]
    \begin{minipage}[t]{0.5\textwidth}
        \centering
        \includegraphics[width=0.9\linewidth]{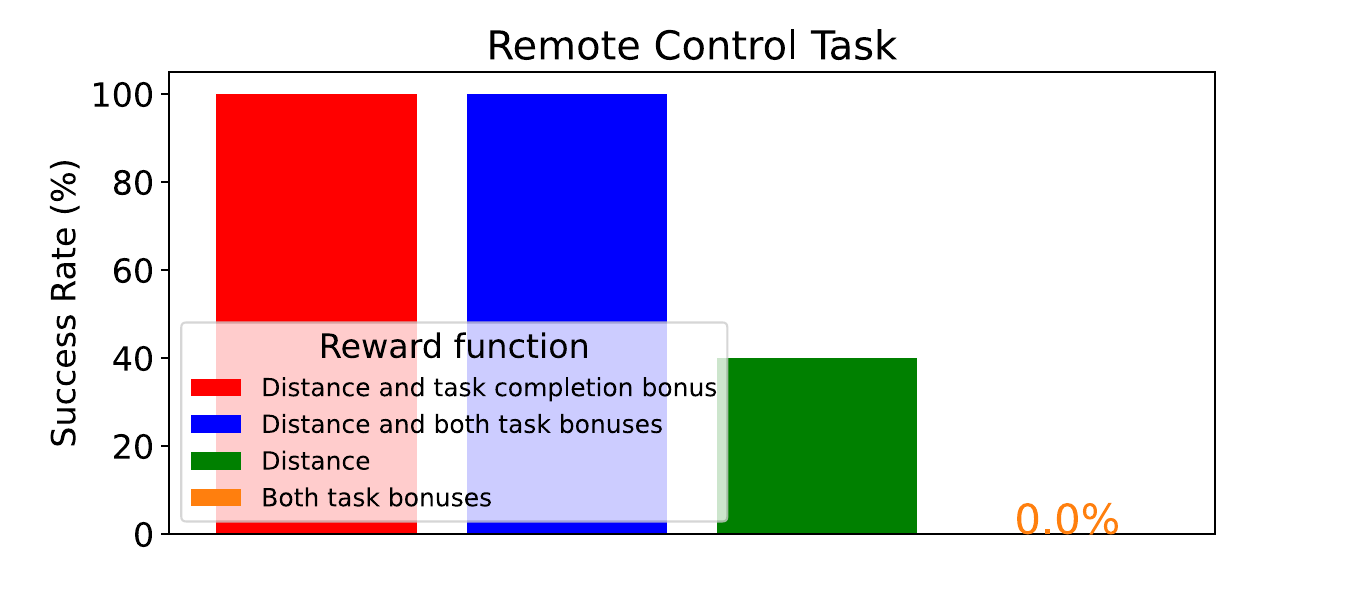}
    \end{minipage}
    \hspace{0.01em}
    \begin{minipage}[t]{0.5\textwidth}
        \centering
        \includegraphics[width=0.9\linewidth]{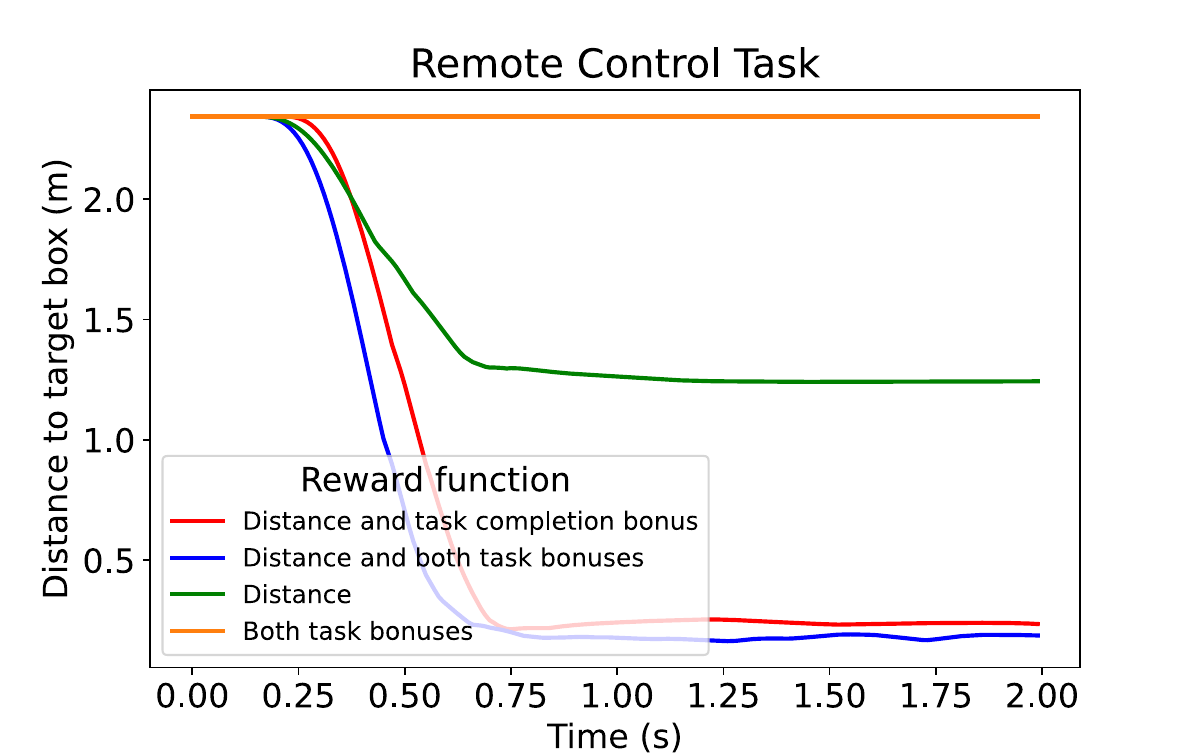}
    \end{minipage}

    \caption{Comparison of success rates (top) and distances to the target (bottom) for policies trained with various reward functions in the remote control task. Combining distance and bonus terms is essential to ensure success. Without a distance term, the car does not move.}
    \label{fig:Remote_control_success_rates}
    \Description{Two plots showing the results of policies trained with the reward functions: Distance and task bonus, Distance and both task bonuses, Distance, and Both task bonuses in the remote control task. The upper plot shows the success rate as percentages. The reward functions consisting of distance and bonus respectively the two bonuses achieve 100\%, distance-only achieves 40 \% and bonus-only 0\%.
    The bottom plot shows the distances (0-2.5 meters) to the target box in meters over time (10 seconds).Policies using distance and task bonus or Distance and both task bonuses quickly (after 4 seconds) reduce the distance to the target to zero. With bonus-only the distance does not change and with distance-only it only reduces to 1.6 meters.}
\end{figure}

\paragraph{Remote Control} We additionally apply our framework to the \textbf{remote control} task and present the results in Figure~\ref{fig:Remote_control_success_rates}. Each agent is tested over 10 episodes. 
Following our guidelines, we again start with a reward function consisting of a distance term and a task bonus (G1, G2). 
The task requires a sequence of actions: moving the fingertip to a joystick and using it to steer a car into a target box. Accordingly, we include two distance terms in the reward function: one penalizing the distance between the fingertip and the joystick, and one penalizing the distance between the car and the target box (G1).
For both distance terms, we use the exponential distance. %
Since the task only requires the car to reach the target box, but allows for further corrections ("parking") after the time limit is reached (i.e., no early termination is applied)~\cite{ikkala_breathing_2022}, a success bonus for reaching the target box can be given for multiple timesteps. %
Therefore, we choose a small bonus weight $w_\text{bonus} = 1$, following G3. %
The resulting agent achieves a 100\% success rate, consistently steering the car into the target. 

Since this task is sequential, one might also consider joystick contact as an intermediate goal state required for correct task execution (G2). An agent trained with this additional, second bonus term also achieves 100\% success and exhibits slightly faster car movement toward the target.
Further results reconfirm the necessity of both distance guidance and a task bonus (G1, G2). The agent trained using only distance costs succeeds in only 40\% of the episodes. Failed episodes result from the agent's inability to reliably control the joystick to steer the car into the target box. %
The agent trained with task bonuses only fails completely, resulting in a 0\% success rate.
It either performs arbitrary movements or does not move at all, demonstrating the importance of guidance in the reward function. Due to the high computational cost of this task and the success of our baseline reward design, we do not train policies with additional effort models for this case (G4).

\section{Discussion}
We here summarize and contextualize our findings with respect to the research questions (Sections~\ref{subsec:disc-plausibility}-\ref{subsec:disc-generalizability}) and identify limitations and avenues for future work (Section~\ref{subsec:disc-limitations}).

\subsection{RQ.1: Plausibility}\label{subsec:disc-plausibility}
Our analyses  identify a dense distance term as an essential reward component to produce plausible, aimed movements across the five considered tasks.
This distance term 'guides' the simulated user towards the desired target, e.g., the correct button in a keyboard typing task or the controller in an instrumented interaction task.

Many HCI tasks impose additional constraints beyond simple distance minimization, such as dwell time, force control, object grasping, or discrete gesture execution.
This leads to a division of the task into at least two phases: (1) a reaching/aiming phase, where distance-based rewards are usually sufficient; and (2) a completion phase. %
For (2), additional reward terms are required. 
We found that providing a bonus term once upon task completion to be sufficient for the considered tasks, e.g., to learn keeping the fingertip inside the target sphere for a fixed dwell time (pointing), or applying enough force for a successful button pressing (choice reaction).
Depending on the time and visuomotor complexity of the completion phase, we expect continuous reward terms, which, e.g., incentivize increasing the button pressing force, to provide additional guidance and thus simplify training. However, further analysis of such constraint-specific guidance terms is left for future work. %
In the tracking task, which simply requires keeping the fingertip close to the moving target, the bonus term proved mostly unnecessary when combined with a guiding distance. 
Rather, the bonus available in every time step plays a ``guiding'' role similar to the distance term, which reinforces the need to balance the two.

Our observations suggest that an effort model is not necessary for successful task completion. 
While biomechanical simulations traditionally use an effort model to constrain the overactuated action space and regularize the optimization landscape~\cite{guigon_computational_2007, berret_evidence_2011, charaja_generating_2024}, our observations do confirm earlier findings from~\cite{fischer2021reinforcement}.
In this work, a biomechanical arm model trained without an effort model was shown to generate mid-air movements that follow well-established characteristics such as Fitts' Law~\cite{mackenzie1992fitts} and the 2/3 Power Law~\cite{schaal2001origins}.

\new{Based on these results, we propose a set of reward design guidelines that are meant to be applied sequentially. While trade-offs, such as between distance and continuous bonus terms, may occur, they can be effectively balanced by following Guideline G3.}

\subsection{RQ.2: Sensitivity}\label{subsec:disc-sensitivity}
While we found that the weighting of the reward function components affected the quality of the trained agents, the performance metrics and the observed qualitative behavior were quite robust to weight changes.
In the choice reaction task, any bonus weight between $0.5$ and $100$ resulted in $100\%$ accuracy when trained without an effort model, albeit with slower training times for larger bonus weight. Similarly, hardly any differences were observed between the tested distance weights, which spanned three orders of magnitude, when effort models were excluded.
However, it is important to strike a good balance between effort and distance weights. %
Effort weights should not be chosen too large to avoid task failure, while the range of 'working' weights depends on the specific effort model. The \ref{DC} effort model that directly penalizes squared muscle controls proved most robust against changes in the weighting.
In the tracking task, the bonus term is not given only once per trial, but can be obtained multiple times, resulting in a larger accumulated sum of these reward terms over a trial. We hypothesize that this is the reason for the observed advantage of smaller bonus weights. However, any bonus weight between $0.125$ and $1$ resulted in good performance and plausible movement.
Effects of distance and effort weights were similar to the choice reaction task.

Our findings suggest that designing reward functions does not require finding a very specific, unique set of cost weights, but rather allows for a fairly wide range of combinations that HCI researchers and designers might come up with, as long as some basic principles are followed.
\new{Adhering to the proposed guidelines promotes a more principled design process, which replaces prevailing unstructured trial-and-error approaches. While a few iterations of adjusting weights within Guidelines G3 and G4 might still be required, these guidelines provide clear instructions how to systematically proceed.} %

\subsection{RQ.3: Generalizability}\label{subsec:disc-generalizability}

Based on the insights gained from over 450 policies trained in the choice reaction, pointing, and tracking task, we were able to distill a set of concise and clear guidelines for reward design.
\new{The analysis focused on three reward components chosen for their relevance and demonstrated effectiveness in goal-directed visuomotor tasks: (1) task completion bonuses, which are standard in RL; (2) proximity rewards, which help guide agents when task success is sparse; and (3) effort terms, which support human-plausible behavior as shown in prior HCI and motor control research~\cite{berret_evidence_2011, charaja_generating_2024}.}
The resulting guidelines propose a sequence of actionable steps that address HCI researchers and designers with little or none experience, and may help them developing their own task-specific biomechanical simulations.
We have validated our guidelines on two independent standard interaction tasks (keyboard typing and remote control), demonstrating their applicability across diverse goal-directed visuomotor tasks.

While the five tasks considered cover a wide range of motor requirements, success conditions and contexts, there are many more interesting use cases that could benefit from biomechanical testing.
We believe that the proposed guidelines can inspire the community to create simulated users for their own interaction techniques, task requirements, and user models.
Extending the scope of RL-based biomechanical simulations is essential to consolidate and deepen our understanding of the role of composite reward functions. Such future work can ultimately form the basis for ongoing evaluation and refinement of the proposed guidelines, which we hope will catalyze this research activity.
\new{In addition, it would be interesting to see how %
advanced RL training methods, such as adaptive curricula~\cite{narvekar2020curriculum, chiappa2024acquiring} and muscle-specific exploration strategies~\cite{schumacher_dep-rl_2022, berg2024sar, chiappa2023latent} %
complement reward design. These approaches could improve learning efficiency and robustness, but they also increase the complexity of the training process.}

\subsection{Limitations and Future Work}\label{subsec:disc-limitations}

While our reward function design provides a structured approach for designing reward functions for RL agents in HCI tasks, several limitations remain. 

In our analysis, we have only considered task bonus terms that were awarded for \textit{successful} task performance.
In environments that enable unsuccessful early termination (e.g., when a walking agent falls down), penalties for undesirable or failure states may help shape agent behavior. Integrating such penalties is straightforward within our framework, but requires further investigations.

\new{Similarly, this work focused on distance terms as the main guiding concept. Extensions to higher-order reference signals, such as maintaining a fixed target speed, matching orientations in 6 DoF docking tasks, or tracking a reference velocity for virtual world navigation are conceivable. Moreover, additional reward components incentivizing aspects beyond success criteria such as ergonomics (e.g. reducing discomfort or pain), novelty or multi-agent collaboration could be integrated. The interplay between these additional factors and the three core reward components considered in this paper warrants further investigation.} %

Future work should also test our guidelines for a wider range of biomechanical models, e.g., including head or eye movements or actuable fingers, \new{and develop more quantitative metrics to assess the motion plausibility of predicted movements and reduce reliance on manual inspection of movement videos prone to subjectivity.} %
In addition, potential biases introduced by shaped reward terms, such as favoring certain targets or body postures, require further investigation to ensure that learned behaviors remain natural and unbiased. This is particularly relevant for modeling more strategic or creative interaction tasks. %

Finally, modular simulation designs that separate the perceptual, cognitive, and motor control tasks that are currently intertwined in our end-to-end visuomotor learning approach will allow for better interpretability of individual reward design choices and more targeted diagnosis of learned behaviour.

\section{Conclusion}

Reward function design is critical for RL-based biomechanical simulations. We systematically investigated the influence of effort minimization, task bonuses, and target proximity incentives across representative HCI tasks: pointing, tracking, and choice reaction. We then performed an extensive analysis of how sensitive task success and completion time depend on the weights of these three reward components. 
Our analysis revealed that a guidance reward component is essential for task completion, while task bonuses are necessary for tasks with additional constraints beyond simple distance minimization. 
Notably, agents can succeed without an effort model, and the addition of effort models requires careful tuning to avoid performance degradation.
Building on these findings, we proposed practical guidelines for reward function design that enable plausible, biomechanical simulations without requiring much RL expertise. We validated these guidelines on two additional tasks: remote control and keyboard typing. %
This work highlights the need for a deeper understanding of the subtleties involved in training RL-based user models, and contributes to making biomechanical user simulations a viable, low-barrier tool for HCI research and design.

\begin{acks}
 This work was supported by EPSRC grant EP/W02456X/1.
 Hannah Selder and Arthur Fleig acknowledge the financial support by the Federal Ministry of Education and Research of Germany and by Sächsische Staatsministerium für Wissenschaft, Kultur und Tourismus in the programme Center of Excellence for AI-research „Center for Scalable Data Analytics and Artificial Intelligence Dresden/Leipzig“, project identification number: ScaDS.AI.
\end{acks}

\bibliographystyle{ACM-Reference-Format}
\bibliography{sample-base}

\end{document}